\documentclass[a4paper,12pt]{article}
\def\al{\alpha}
\def\be{\beta}
\def\ga{\gamma} \def\Ga{\Gamma}
\def\ep{\epsilon}
\def\lam{\lambda}
\def\Lam{\Lambda}

\def\calA{{\cal A}} \def\calB{{\cal B}}  
  
 \def\calH{{\cal H}} 
  
 \def\calN{{\cal N}} \def\calO{{\cal O}}
\def\calP{{\cal P}}  \def\calR{{\cal R}}
\def\calS{{\cal S}}

\def\del        {  \partial  }
\def\half       {  {1\over 2}  }

\def\ie         {  {\it i.e.}      }

\def\comma          {\, ,}
\def\period         {\, .}
\def\lsim    {\lower .65ex \hbox{\ $\stackrel{<}{\sim}$\ } }
\def\gsim    {\lower .65ex \hbox{\ $\stackrel{>}{\sim}$\ } }
\def\com#1#2   { \left[#1, #2\right]} 
\def\acom#1#2  {\left\{ #1,#2\right\}}
\def\bra#1     {\langle #1 |}
\def\ket#1     {| #1 \rangle}
 
\def\vecii#1#2      {  \left(\begin{array}{c}#1\\#2\end{array}\right)  }
\def\veciii#1#2#3   {  \left(\begin{array}{c}#1\\#2\\#3\end{array}
                     \right)  }
\def\veciv#1#2#3#4  {  \left(\begin{array}{c}#1\\#2\\#3\\#4
                                 \end{array}\right)  }
\def\vecfv#1#2#3#4#5 {  \left(\begin{array}{c}#1\\#2\\#3\\#4\\#5
                                 \end{array}\right)  }

\def\matrixii#1#2#3#4            {  \left(\begin{array}{cc}#1&#2\\#3&#4
                                       \end{array}\right) }
\def\matrixiii#1#2#3#4#5#6#7#8#9 {  \left(\begin{array}{ccc}#1&#2&#3\\
                                     #4&#5&#6\\#7&#8&#9\end{array}
                               \right)  }
\def\mativ#1#2#3#4               {  \left(\begin{array}{cccc}
                                       #1\\#2\\#3\\#4\end{array}\right) }
\def\matv#1#2#3#4#5              {  \left(\begin{array}{ccccc}
                                     #1\\#2\\#3\\#4\\#5\end{array}
                              \right)  }
\def\eqabegin         {  \begin{eqnarray}  }
\def\eqaend           {  \end{eqnarray}  }
\def\nn               {  \nonumber  }
\def\bracetwo#1#2     {  \left\{ \begin{array}{l} #1 \\ #2 \end{array}
                         \right.  }
\def\bracetwocases#1#2#3#4  {   \left\{ \begin{array}{ll} #1 &
                                 \qquad #2 \\
                                 #3 & \qquad #4 \end{array} \right.  }
\def\bracebegin#1     {  \left\{ \begin{array}{#1}   }
\def\braceend         {  \end{array}\right.   }
\def\parn              {  \par\noindent }

\def\parmedskip        {  \par\medskip  }

\def\parag#1           {\paragraph{#1} \mbox{ }\parmedskip\noindent}
\def\msection#1      {  \begin{center} \section{#1} \end{center}   }
\def\nsection#1      {  \let\boldface\bf \def\bf{} \section{#1}
                           \let\bf\boldface   }
\def\mnsection#1     {  \begin{center} \nsection{#1} \end{center}  }
\def\capsection#1    {  \let\boldface\bf \def\bf{\sc} \section{#1}
                           \let\bf\boldface   }
\def\mcapsection#1   {  \begin{center} \capsection{#1} \end{center} }

\newcommand{\nullify}[1]{}
\def\papertitlepage{\baselineskip 3.5ex \thispagestyle{empty}}
\def\Title#1{\baselineskip 1cm \vspace{1.5cm}\begin{center}
 {\Large\bf #1} \end{center} 
\vspace{0.5cm}}
\def\Authors#1{\begin{center} {\it #1} \end{center}}
\def\Abstract{\vspace{1.0cm}\begin{center} {\large\bf Abstract} 
           \end{center} \par\bigskip}
\def\Komabanumber#1#2#3{\hfill \begin{minipage}{4.2cm} UT-Komaba #1
              \parn #2 
              \parn #3 \end{minipage}}
\renewcommand{\thefootnote}{\fnsymbol{footnote}}
\renewenvironment{thebibliography}{\pagebreak[3]\par\vspace{0.6em}
\begin{flushleft}{\large \bf References}\end{flushleft}
\vspace{-1.0em}

\begin{enumerate}\if@twocolumn\baselineskip=0.6em\itemsep -0.2em
\else\itemsep -0.2em\fi\labelsep 0.1em}{\end{enumerate}}

\usepackage{graphicx}
\usepackage{latexsym}
\usepackage{amsmath}
\usepackage{amssymb}

\renewenvironment{thebibliography}{\pagebreak[3]\par\vspace{0.6em}
\begin{flushleft}{\large \bf References}\end{flushleft}
\vspace{-1.0em}

\begin{enumerate}\if@twocolumn\baselineskip=0.6em\itemsep -0.2em
\else\itemsep -0.2em\fi\labelsep 0.1em}{\end{enumerate} }
\def\Gabar{\bar{\Ga}}

\def\Gatil{{\tilde{\Ga}}}

\def\bk#1#2{{\langle #1 | #2 \rangle }}

\def\Itil{{\tilde{I}}}
\def\Jtil{{\tilde{J}}}
\def\itil{{\tilde{i}}}

\def\up{u^+}
\def\um{u^-}

\def\Qtil{{\tilde{Q}}}

\def\gcom#1#2{{\left[ #1, #2\right\}}}

\def\Qhat{\hat{Q}}
\def\Fmn{F_{\mu\nu}}

\def\calNbar{\bar{\calN} }
\def\calNtil{\widetilde{\calN}}
\def\lampinv{\lam_+^{-1}}
\def\Abar{{\bar{A}}}

\def\Llam{L_{(\lam)}}

\def\Vtil{\tilde{V}}

\def\Atil{\tilde{A}}
\def\Btil{\tilde{B}}
\def\npb#1{Nucl. Phys. {\bf B#1}}
\def\prd#1{Phys. Rev. {\bf D#1}}
\def\jhep#1{JHEP {\bf #1}}
\def\hepth#1{ hep-th/#1}
\def\plb#1{Phys. Lett. {\bf B#1}}

\begin{document}
\papertitlepage
\vspace*{0cm}
\Komabanumber{02-15}{hep-th/0212316} {December, 2002}
\Title{A New First Class Algebra, Homological Perturbation and 
Extension of Pure Spinor Formalism  for Superstring} 
\Authors{{\sc Yuri Aisaka\footnote[2]{yuri@hep1.c.u-tokyo.ac.jp} 
 and Yoichi Kazama
\footnote[3]{kazama@hep1.c.u-tokyo.ac.jp}
\\ }
\vskip 3ex
 Institute of Physics, University of Tokyo, \\
 Komaba, Meguro-ku, Tokyo 153-8902 Japan \\
  }
\baselineskip .7cm

\numberwithin{equation}{section}
\numberwithin{figure}{section}
\numberwithin{table}{section}

\parskip=0.9ex

\Abstract

Based on a novel first class algebra, we develop an extension of 
the pure spinor (PS) formalism of Berkovits, in which the PS constraints
are removed. By using the  homological perturbation theory in 
an essential way, the BRST-like
charge $Q$ of the conventional PS formalism is promoted to a bona fide 
nilpotent charge $\hat{Q}$, the cohomology of which is equivalent 
to the constrained cohomology of $Q$. This construction requires 
only a minimum number (five) of additional fermionic ghost-antighost pairs and 
the vertex operators for the massless modes of open string are obtained 
in a systematic way. Furthermore, we present a simple 
composite ``$b$-ghost" field $B(z)$ which realizes the important relation 
$T(z) = \{ \hat{Q} , B(z)\} $, with $T(z)$ the Virasoro operator, and 
apply it to facilitate the construction of the integrated vertex. 
The present formalism utilizes  $U(5)$ parametrization
and the manifest Lorentz covariance is yet to be achieved.

\newpage
\baselineskip 3.5ex
\section{Introduction}  
\renewcommand{\thefootnote}{\arabic{footnote}}
Desires to construct a quantization scheme for superstring in which 
 both the Lorentz symmetry  and the spacetime supersymmetry 
are manifest are ever mounting in the recent striking 
developments of string theory. Apart from the obvious aesthetic appeal,
 it should be indispensable for deeper understanding of various issues, 
such as the S-duality of type IIB theory and multitude of important
 problems involving Ramond-Ramond fields. 

The conventional Ramond-Neveu-Schwarz (RNS)\cite{RNS} and Green-Schwarz (GS)
\cite{GS81} formalisms are well-known to be inadequate for this purpose. 
In RNS formalism, the Lorentz symmetry is manifest but the spacetime 
 supersymmetry is not, requiring GSO projection for its implementation.
 In contrast, GS 
 formalism, which is capable of realizing 
 both of these symmetries at the classical  level, 
is nevertheless difficult to quantize  except in the light-cone type 
non-covariant gauge. 

About three years ago, following earlier 
attempts \cite{Siegel86,BVparticles,HSstring,BVparticlesOK,TWstring,Hybrid}, 
Berkovits initiated a new formalism\cite{Berk0001} based 
 on the concept of pure spinor (PS)\cite{Cartan, Howe91a, Howe91b}, 
 which is super-Poincar\'e covariant and posesses a number of 
remarkable features. 
The formalism is based on free fields, which form a conformal field 
 theory (CFT) with vanishing central charge.
 The principal ingredient  of the formalism is the BRST-like charge 
$Q=\int dz \lam^\al d_\al$, 
 where $d_\al$ is the spinor covariant derivative and $\lam^\al$ 
 is a bosonic chiral spinor field satisfying the pure spinor 
 constraints $\lam^\al \ga^\mu_{\al\be} \lam^\be =0$. 
With these  constraints $Q$ becomes  nilpotent and  inherently second class 
constraints $d_\al$ can be consistently implemented as $Q\psi=0$ without 
losing manifest Lorentz covariance. All the perturbative 
physical states of superstring have been demonstrated to be realized 
 precisely as the elements of the constrained cohomology of $Q$ 
\cite{Berk0006}. Spacetime supersymmetry is manifestly maintained 
throughout the formalism without the need of GSO projection nor the 
 complication due to picture changing. Covariant rules can be 
 given with which one can compute the scattering amplitudes 
 in a manifestly super-Poincar\'e covariant manner\cite{Berk0001, 
Berk0004}. Although rather involved, the relation to RNS formalism 
 has essentially been understood \cite{Berk0104}. This formalism 
 can be easily applied to a superparticle\cite{Berk01050}
 and an intriguing application 
 to supermembrane has also been suggested\cite{Berk0201}. 
Other recent developments  are found in \cite{PSrecent}. 
The rudiments
 of this  formalism will be briefly reviewed in Sec.2, 
while for more comprehensive review we refer the reader to \cite{Berk0209}. 

Quite attractive as it is, there are many challenges for PS formalism. 
Essentially  they all stem from the lack of understanding of the 
  origin of the formalism. The underlying symmetry principle and the 
 fundamental action are not known. This is reflected on the as yet obscure 
 nature of the BRST-like charge, although one can relate it to RNS 
 formalism in a round-about way\cite{Berk0104}. In this regard, recently 
 an attempt has been made to derive the Berkovits formalism 
from the superembedding approach\cite{Matone0206}. This is an interesting and 
 aspiring endeavor but so far the analysis is completely classical and the 
justification of the identification of the BRST charge is not yet achieved. 

Continuing along the list of to-be-improved items, we must mention 
the  covariant rules proposed for calculations of 
scattering amplitudes. These rules are well-motivated and 
 reasonable but they should be derived from more basic principles. 
In fact, surprisingly, to our knowledge a serious study of the inner 
 product structure of the PS formalism has not been performed. 
An analysis of this issue will be presented in Sec.3, which will 
point to another important question about PS formalism, namely 
 whether PS constraints are really necessary. One would like the answer
 to be \lq\lq no"  as the removal of the PS constraints 
 should  be more natural and useful in the quantization of eventual 
 underlying action. Recently, a proposal in this direction was made in 
\cite{stonybrook}, where a BRST-like nilpotent charge was 
 constructed without PS constraints in a step-wise covariant manner 
based on a Kac-Moody algebra with a central charge.
 This formalism has many desired properties 
 but to get non-trivial cohomology one must impose
 an extra condition and this makes the formalism rather involved.  
Also an attempt to remove the PS constraints for the superparticle case 
 has recently been reported \cite{Chesterman}. The proposed scheme is 
 covariant and shares some features with our present work but appears to suffer
 from the long-standing problem of ghosts for ghosts ad infinitum. 

In this paper, we shall develop a different scheme to remove 
 the PS constraints from the Berkovits' formalism. Our formalism 
 is based on a novel closed {\it first class algebra} \ie without 
 central charges, a member of which 
 is the BRST-like current $j(z)$ of Berkovits but now without the 
PS constraints. Formulation of this algebra and the subsequent 
construction of a nilpotent charge and vertex operators are
 performed in the so-called $U(5)$ parametrization,
 as in the original PS formalism, so that these intermediate steps 
 are not manifestly Lorentz covariant. This is certainly a disadvantage 
 compared to the approach of \cite{stonybrook} but we gain considerably 
 in the simplicity of the formalism. As it will be explained in 
detail in Sec.5, the construction of a new nilpotent BRST-like charge, 
to be called $\Qhat$,  requires only a minimum number  
 of anti-commuting ghost pairs $(c_I,b_I)_{I=1\sim 5}$  and  turned out 
 to fit beautifully 
into the scheme of homological perturbation theory\footnote{This 
 scheme has been used in \cite{Berk0104} in a step relating 
 PS to  RNS formalisms.}\cite{HenTb} in a profound way. 
The biggest advantage of this scheme is that the cohomology of $\Qhat$ 
 is easily shown to be exactly equivalent to the constrained cohomology of 
$Q$. Furthermore, the logic of the proof of this equivalence 
 can be applied to the systemtic construction of the vertex operators,
 both unintegrated and integrated, in a transparent fashion. 
Another important outcome of our investigation is that, in the course 
 of the analysis of the integrated vertex operator, we discovered 
a remarkably simple composite \lq\lq $b$-ghost" field $B(z)$, 
which realizes the fundamental relation 
$T(z) = \{ \hat{Q} , B(z)\} $, where $T(z)$ is the Virasoro operator
of our system. This provides an alternative far simpler method of 
 construction for the integrated vertex. This relation to the 
 Virasoro generator, which hitherto has been rather elusive in PS formalism,
is known to be of prime importance for no-ghost theorem and loop calculations 
\cite{FreemanOlive} and is expected to play crucial roles in future 
 developments.

We organize the rest of this paper as follows:\ In Sec.~2, we start with  a 
 short review of  Berkovits' pure spinor formalism, which at the 
 same time introduces our notations and conventions. 
In Sec.~3, we present an analysis of the appropriate inner product 
 in the PS formalism,
with which one should  implement the peculiar hermiticity property 
 of the pure spinor. This study leads to yet another motivation for 
removing the  PS constraints. The main results of this paper 
 will be described starting from Sec.~4. In Sec.4, we construct 
 a new closed first class algebra,  which will be the basis of our 
 formalism. Then, in Sec.~5, after giving a brief description of 
 the construction of a nilpotent BRST-like charge via Batalin-Vilkovisky 
 procedure, we apply an alternative scheme of
homological perturbation theory, more natural and far-reaching 
 in the present context,  and obtain a simpler nilpotent 
 charge $\Qhat$. Subsequently, we sketch 
the proof of the equivalence of the $\Qhat$-cohomology to that of the 
 constrainted cohomology of $Q$ in the Berkovits' formalism. Having 
 understood all the necessary fields of our formalism, we end this 
 section  by displaying the free action and the energy-momentum tensor. 
Sec.~6 will be devoted to the systematic construction of 
 the physical vertex operators for the massless modes of open 
superstring. Both the 
 unintegrated and the integrated vertices are obtainted in a systematic 
 manner. Further, we present the simple composite 
\lq\lq $b$-ghost" field $B(z)$, prove the key relation 
$T(z) = \{ \hat{Q} , B(z)\} $ and give an alternative more superior 
 derivation of the integrated vertex. 
Finally in Sec.~7, we summarize our findings and discuss remaining issues. 
Two appendices are provided to explain some technicalities: 
In Appendix A, we summarize our conventions for $\Ga$-matrices and 
$U(5)$ parametrizations. 
Appendix B is devoted to the proof of the triviality of the $\delta$-homology, 
which forms the basis of the homological perturbation. 
\section{Rudiments of Pure Spinor Formalism}
We begin with a brief review of the essential ingredients of the pure spinor 
 formalism, which at the same time serves to introduce our notations and 
 conventions. Throughout we refer only to the holomorphic sector, 
 appropriate for open string. 

The central idea of the pure spinor formalism proposed by Berkovits 
\cite{Berk0001} is that the physical states of superstring 
can be described as the elements of the cohomology of a BRST-like 
 operator $Q$ given by\footnote{For simplicity we will use 
 the notation $[dz] \equiv dz/(2\pi i)$ throughout.}
\begin{eqnarray}
Q &=& \int [dz] \lam^\al(z) d_\al(z) \comma \label{defQ}
\end{eqnarray}
where $\lam^\al$ is a $16$-component bosonic chiral spinor 
 satisfying the  pure spinor constraints 
\begin{eqnarray}
\lam^\al \ga^\mu_{\al\be} \lam^\be &=& 0 \comma \label{psconst}
\end{eqnarray}
and $d_\al$ is the spinor covariant derivative given in our 
convention\footnote{Our convention, including  normalization, 
 of  a number of  quantities are slightly different from those often (but 
 not invariably) used by 
 Berkovits. Our convention has been chosen to make the description 
 more standard and hence familiar to non-experts in this field. 
Some further details of our convention are described  in 
 Appendix A. }
by 
\begin{eqnarray}
d_\al &=& p_\al + i\del x_\mu (\ga^\mu\theta)_\al + \half 
(\ga^\mu\theta)_\al (\theta \ga_\mu \del \theta) \period \label{defdal}
\end{eqnarray}
Here $x^\mu$ and $\theta^\al$ are, respectively, the basic bosonic and 
ferminonic worldsheet fields describing a superstring, which transform 
 under the spacetime supersymmetry with global spinor parameter $\ep^\al$ 
 as 
\begin{eqnarray}
\delta \theta^\al &=& \ep^\al \comma \qquad 
\delta x^\mu = i\ep \ga^\mu \theta\period \label{susytransf}
\end{eqnarray}
$x^\mu$ is self-conjugate and satisfies  $x^\mu(z) x^\nu(w) 
 = -\eta^{\mu\nu} \ln (z-w)$, while $p_\al$ serves as the conjugate 
 to $\theta^\al$ in the manner $\theta^\al(z) p_\be(w) 
 = \delta^\al_\be /(z-w)$. $\theta^\al$ and $p_\al$ carry 
 conformal weights 0 and 1 respectively. With such free field operator 
 product expansions (OPE's), 
$d_\al$ satisfies the following OPE with itself:
\begin{eqnarray}
d_\al (z) d_\be(w) &= & {2i \ga^\mu_{\al\be} \Pi_\mu(w) 
\over z-w} \comma \label{ddope}
\end{eqnarray}
where $\Pi_\mu$ is the basic superinvariant combination 
\begin{eqnarray}
\Pi_\mu &=& \del x_\mu -i\theta \ga_\mu \del \theta \period 
\label{defPi}
\end{eqnarray}
Then, due to the pure spinor constraints (\ref{psconst}), $Q$ is easily 
 found to be nilpotent and the  constrained cohomology of $Q$ 
 can be defined. The basic superinvariants $d_\al, \Pi^\mu$ and 
$\del\theta^\al$  form the closed algebra 
\begin{eqnarray}
d_\al(z) d_\be(w) &=& {2i\ga^\mu_{\al\be} \Pi_\mu (w)\over z-w} 
\comma \label{dd2}\\
d_\al(z) \Pi^\mu(w) &=& {-2i(\ga^\mu \del\theta)_\al(w) \over z-w} 
\comma \label{dPi2}\\
\Pi^\mu(z) \Pi^\nu(w) &=& -{\eta^{\mu\nu} \over (z-w)^2} 
\comma \label{PiPi2}\\
d_\al(z) \del \theta^\be(w) &=& {\delta^\be_\al \over (z-w)^2} 
\comma \label{ddelth2}
\end{eqnarray}
which has a central charge and hence is essentially of second class. 
The supersymmetry transformation is generated by the 
supercharge\cite{Siegel86} 
\begin{eqnarray}
q_\al &=& \int [dz] \left\{ p_\al - i\del x_\mu (\ga^\mu\theta)_\al 
- {1\over 6}
(\ga^\mu\theta)_\al (\theta \ga_\mu \del \theta) \right\}
\comma  \label{susycharge}
\end{eqnarray}
which obeys the supersymmetry 
algebra $\acom{q_\al}{q_\be} = -2i \ga^\mu_{\al\be}
\int[dz] \Pi_\mu(z)$. 

Although eventually all the rules can be formulated in a Lorentz 
covariant manner, various quantities involving the pure spinor $\lam$ 
 are first defined transparently in the so-called  $U(5)$ basis\footnote{Our 
 conventions for $U(5)$ parametrization is summarized in Appendix A}. 
As is well-known, 
 the spinor representations for $SO(9,1)$ and $SO(10)$ can be 
efficiently constructed using 5 pairs of fermionic oscillator variables 
 $(b_I, b_I^\dagger)$ satisfying
 $\acom{b_I}{b_J^\dagger} =\delta_{IJ}$, which in the case of $SO(10)$ 
 transform as $(5,\bar{5})$ of $U(5)$ subgroup\footnote{When we discuss
 the hermiticity property of $\lam$ in Sec.3, the distinction 
 between $SO(9,1)$ and $SO(10)$ becomes important. 
 However, since the conversion between them is rather trivial
 for other purposes, we will use the terminology appropriate for 
$SO(10)$.}
. In this basis, to be often 
 referred to as a {\it U-basis} in this work, a chiral spinor $\lam$
 is described by the component fields  $\lam_A$ given by 
\begin{eqnarray}
\lam_A &=& (\lam_+, \lam_{IJ}, \lam_\Itil) \sim  (1,10,\bar{5}) \in 
 U(5) \comma   \label{lamA}
\end{eqnarray}
where we have indicated how they transform under $U(5)$, with a 
 tilde on the $\bar{5}$ components\footnote{In what follows, to avoid 
unnecessary clutter in notations, we omit tilde for $\bar{5}$ indices 
 except when it is absolutely essential. One can easily recover 
 the correct type of indices if needed.}. In this 
representation, the 
 pure spinor constraints (\ref{psconst}) reduce to the
 5 independent conditions
\begin{eqnarray}
\Phi_I \equiv  \lam_+\lam_\Itil -{1\over 8}
\ep_{IJKLM}\lam_{JK}
 \lam_{KL} =0  \period \label{defPhi} 
\end{eqnarray}
Therefore the number of independent components of a pure spinor is 
 11 and together with all the other fields (including the conjugates 
 to the independent components of $\lam$) the entire system constitutes 
 a free CFT with vanshing central charge. 

The fact that the constrained cohomology of $Q$ is in one to one 
 correspondence with the light-cone degrees of freedom of superstring 
 was shown in \cite{Berk0006} using the $SO(8)$ parametrization of a
 pure spinor. Besides being non-covariant, 
this parametrization contains redundancy and an infinite
 number of supplimentary ghosts had to be introduced. Nonetheless, 
 subsequently the Lorentz invariance of the cohomology 
was demonstrated in \cite{Berk01051}. 

The great advantage of this formalism is that one can compute the 
 scattering amplitudes under a set of rules which are 
manifestly super-Poincar\'e covariant. For the massless modes 
the physical unintegrated vertex operator is given by a simple form 
\begin{eqnarray}
U_0 &=& \lam^\al A_\al(x,\theta) \comma \label{unintvo}
\end{eqnarray}
where $A_\al$ is a spinor superfield satisfying the \lq\lq on-shell" condition
\begin{eqnarray}
(\ga^{\mu_1\mu_2 \ldots \mu_5})^{\al\be} D_\al A_\be &=& 0\comma 
\label{onshell}
\end{eqnarray}
with 
\begin{eqnarray}
D_\al = {\del\over \del\theta^\al} -i(\ga^\mu\theta)_\al
{\del \over \del x^\mu} 
\period \label{defDal}
\end{eqnarray}
Then, together with the pure spinor constraints, $QU_0=0$ is easily 
 verified and moreover $\delta U_0 =Q\Lam$ represents the gauge transformation
 of $A_\al$. Its integrated counterpart $\int[dz]V_0(z)$, needed 
 for the calculation of $n$-point amplitudes with $n\ge 4$, is 
 characterized by $QV_0 =\del U_0$ and was constructed 
to be of the form \cite{Berk0001} \cite{Siegel86}
\begin{eqnarray}
V_0 &=& \del\theta^\al A_\al + \Pi^\mu B_\mu + d_\al W^\al 
 + \half \Llam^{\mu\nu} F_{\mu\nu} \period \label{intvo}
\end{eqnarray}
Here, $B_\mu =(i/16) \ga^{\al\be}_\mu D_\al A_\be$ is the gauge superfield, 
 $W^\al =(i/20) (\ga^\mu)^{\al\be}(D_\be B_\mu-\del_\mu A_\be)$ is 
 the gaugino superfield,  $F_{\mu\nu}=\del_\mu B_\nu -\del_\nu B_\mu$ 
 is the field strength superfield and $\Llam^{\mu\nu}$ is the Lorentz generator
 for the pure spinor sector. In this construction various 
relations\cite{onshellcond} following solely from  
 the on-shell condition (\ref{onshell}) play crucial roles. 
They will be displayed later in Sec.~6 when we need them.
 In a similar manner, 
 the vertex operators for the first massive modes have also been 
 constructed recently \cite{Berk0204}. 

With these vertex operators, the scattering amplitude is expressed 
 as 
\begin{eqnarray}
\calA =\langle U_1(z_1)U_2(z_2)U_3(z_3) 
\int [dz_4] V_4(z_4) \cdots \int [dz_N] V_N(z_N) \rangle \comma 
\label{amp}
\end{eqnarray}
 and  can be 
 computed in a covariant manner with certain rules assumed for
 the integration over the zero modes of $\lam^\al$ and $\theta^\al$. 
The proposed prescription enjoys a number of required properties and 
 leads to results which agree with those obtained in the RNS formalism
\cite{Berk0001} \cite{Berk0004} \cite{Berk0104}. 

Sketched above are the basic ingredients of the pure spinor formalism. 
There have been a number of applications of this formalism, including 
 those to a superparticle, a supermembrane, etc. For these and other 
 related developments we refer the reader to a recent review 
by Berkovits \cite{Berk0209}. 
\section{Hermiticity and Inner Product in Pure Spinor Formalism}
As was pointed out in the introduction, one of the 
 important issues in PS formalism is the clarification 
 of the Hilbert space structure, in particular the proper 
 definition of the inner product with which the peculiar 
 hermiticity property of the pure spinor $\lam^\al$ should be realized 
 in a natural manner. In this section, we shall examine this problem 
 in some detail and find that with the PS constraints  implementation 
 of an appropriate inner product is extremely difficult if not impossible. 
This observation strengthens the  motivation to remove the PS
 constraints, which will be achieved in subsequent sections. 
\subsection{Hermiticity of $\lam^\al$}
Let us begin with a description of
 the peculiarity of the hermiticiy 
 property of $\lam^\al$,  which is used to construct 
  the basic BRST-like charge $Q=\int[dz] \lam^\al(z) d_\al(z)$.
Up to a certain point, all the discussions will be
 valid without the PS constraints. 
 As it will be clear, the peculiarity 
 shows up in different guises depending on the spinor basis 
chosen.

 First consider the usual basis, to be called a real-basis or an 
{\it R-basis} for short, in which 
 the $SO(9,1)$ $\ga$-matrices are all real. Since $Q$  must be 
 hermitian and in R-basis  $d_\al(z)$, with $\ga^\mu$'s in it being 
 real, satisfies the property $d_\al(z)^\dagger = z^2 d_\al(z)$
 according to the usual hermiticity property of $x^\mu, p_\al$
 and $\theta^\al$, 
$\lam^\al(z)$ in turn must be hermitian, namely 
\begin{eqnarray}
(\lam^\al(z))^\dagger &=& \lam^\al(z) \period \label{hconjlam}
\end{eqnarray}
On the other hand, we know that $\lam^\al$ must be complex in order 
 to satisfy the pure spinor conditions. This by itself is  of course 
 not inconsistent since a hermitian operator can have complex eigenvalues. 
However, we do anticipate some complications as 
the Hilbert space metric becomes necessarily indefinite and there 
 will be null states. 

Actually, since all the basic definitions were made in $U(5)$ parametrization
 in Berkovits' formalism, it is better to study the problem in 
 U-basis, where it appears in a different way \cite{Berk0104}. To see this in more detail and to understand 
 the nature of the issue, we should go back to the basic definition 
of hermiticity for spinor fields in a general basis. As a textbook matter, 
the charge conjugate of a spinor $\psi$ is defined by 
 $\psi^c = B\psi^\ast$ and $\psi$ is real if $\psi^c =\psi$. Here 
$B$ is the matrix satisfying the properties 
$B \Ga^\mu B^{-1} = {\Ga^\mu}^\ast\comma 
 B^\ast =B^T= B\comma  B^2=1 \comma$
and is related to the charge conjugation matrix $C$ by the relation
 $B=-C\Ga^0$. This means that a hermitian spinor operator $\lam$ should be 
 characterized by 
\begin{eqnarray}
\lam^\dagger &=& B\lam \period \label{defherm}
\end{eqnarray}
This of course reduces to the simple form (\ref{hconjlam}) in an R-basis, 
where $B=1$. In a general basis, however, $B$ cannot be taken to be 
 unity. This is because it does not transform by a similarity transformation
 under the change of spinor basis: Under a basis transformation 
 $\Gatil^\mu = T\Ga^\mu T^{-1}$, $B$ transforms as $\tilde{B}
= TB{T^\ast}^{-1}$ and this is not a similarity transformation unless 
 $T^\ast =T$. 

Let us now go to the U-basis and spell out the condition of hermiticity. 
In this basis, $C$  and $B$ matrices are given by 
$C = -\Ga^0\Ga^2 \Ga^4\Ga^6\Ga^8$, $B = -\Ga^2 \Ga^4\Ga^6\Ga^8$.
To write out the explicit content of (\ref{defherm}), it is convenient 
 to use the representation in terms of the fermionic 
oscillators\footnote{For more details, see Appendix A.} 
$(b_I, b_I^\dagger)$. $B$ is then given by 
$B = -(b_1-b_1^\dagger) (b_2-b_2^\dagger)
(b_3-b_3^\dagger) (b_4-b_4^\dagger)$, 
and satisfies the properties $Bb_0^\dagger =b_0^\dagger B$ and 
 $Bb_i^\dagger =b_i B, (i=1\sim 4)$. Applying $B$ to a chiral spinor 
$\lam$, which in the oscillator representation in U-basis 
can be represented 
by a ket as 
\begin{eqnarray}
\ket{\lam} &=& \lam_+ \ket{+} + \half \lam_{IJ} \ket{IJ} 
 + \lam_\Itil \ket{\Itil} \comma \label{lamket}
\end{eqnarray}
one easily finds that the hermiticity condition (\ref{defherm}) 
amounts to the relations 
\begin{eqnarray}
\lam_+^\dagger &=& -\lam_{\tilde{0}} \comma \nn\\
\lam_{ij}^\dagger &=& \half \ep_{ijkl}\lam_{kl}\comma  \label{lamhermU}\\
\lam_{0i}^\dagger &=& \lam_{\tilde{i}}\period \nn
\end{eqnarray}
Thus, the components of a hermitian chiral spinor in this basis 
 have peculiar properties in that conjugation 
 relates different components. 

It is instructive to display briefly how the components of $\lam$ in 
R- and U- bases are related. In a particular R-basis, for example, 
 components of $\lam^\al$ can be expressed in terms of its U-basis 
 components as 
\begin{eqnarray}
\lam^1 &=& \half (\lam_+ -\lam_{\tilde{0}} -\lam_{23} -\lam_{14})\comma 
 \nn\\
\lam^2 &=& {i\over 2} (\lam_{12}-\lam_{34} +\lam_{13} +\lam_{24})\comma  \nn\\
& \vdots & \nn\\
\lam^{15} &=& {i\over 2} (\lam_{03}-\lam_{\tilde{3}} 
+\lam_{02} -\lam_{\tilde{2}})\comma \nn\\
\lam^{16} &=& \half (\lam_{04}+\lam_{\tilde{4}} 
-\lam_{01}-\lam_{\tilde{1}})\period \nn
\end{eqnarray}
One can easily see that 
 they indeed satisfy ${\lam^\al}^\dagger =\lam^\al$ according to the rules 
(\ref{lamhermU}). 
\subsection{Proper inner product}
Let us now ask how we can realize such a peculiar hermitian conjugation 
property  in the Hilbert space of our CFT. 
In what follows, the usual Fock space 
 inner product will be denoted by $(u,v)$ and the hermitian conjugation with 
respect to it by $u^\ast$. 
In a basis with non-trivial $B$, 
such as in a U-basis, 
 the ordinary Fock space conjugation for the Fourier mode, like 
 ${\lam_n}^\ast =\lam_{-n}$, is clearly insufficient. To remedy this, 
we  introduce 
 a new inner product $\langle u, v \rangle$ by 
\begin{eqnarray}
\langle u, v \rangle &\equiv & (u, S v) \comma 
\end{eqnarray}
where $S$ is a kind of  \lq\lq metric " operator.  
Then, the hermitian conjugation of an operator $\calO$ 
with respect to this new inner product is given   by
$ \langle u, \calO v \rangle = \langle \calO^\dagger u, v\rangle $, 
which  in the Fock space language reads $(u,S\calO v) = 
(\calO^\dagger u, Sv)$. Now 
using the Fock space conjugation,  $(u, S\calO v)$ can be rewritten as
$ ((S^{-1})^\ast \calO^\ast S^\ast u, S v)$, 
and hence we have 
\begin{eqnarray}
 \calO^\dagger &=& (S^{-1})^\ast \calO^\ast S^\ast \period \label{herconj}
\end{eqnarray}
We see that the new conjugation is supplemented with
 a similarity transformation 
 by $S^\ast$ and can be quite non-trivial. 
Since $S$ must be invertible it will be convenient  to write it as
\begin{eqnarray}
S =e^{-R^\ast} \period 
\end{eqnarray}
The new hermitian conjugation on $\calO$ is involutive provided
 $(S^{-1} S^\ast)^{-1} \calO 
 (S^{-1} S^\ast) =\calO$. This is satisfied if $S^\ast =S$ but 
it is not a necessary condition, as we shall see. 
\subsection{Explicit construction of $S$ 
and problem with pure spinor constraints}
Hereafter we will work in a U-basis,  where all the basic properties of 
 the operators in pure spinor formalism have actually been derived. 
We regard the components $\lam_A$ of $\lam$  {\it in this basis}
 to be the basic 
 conformal fields satisfying the usual Fock space hermiticity $\lam_A^\ast
 =\lam_A$. 

Our task now is to construct the operator $R$, which according to
 (\ref{herconj}) effects 
\begin{eqnarray}
\lam^\dagger &=& e^{R} \lam\, e^{-R} = B\lam 
\period \label{lamdag}
\end{eqnarray}
Since $B$ is real, we have $(\lam^\dagger)^\dagger 
 = e^R B\lam e^{-R} = BB \lam =\lam$, so this conjugation is involutive. 
It is useful to note that $B$ can be written as 
\begin{eqnarray}
B &=& e^{-(i\pi/2) (1-B)} \comma \label{Bexp}
\end{eqnarray}
which follows from the property $B^2=1$. Then, it is 
 easy to see that (\ref{lamdag}) is fulfilled 
 if $R$ satisfies 
\begin{eqnarray}
\com{R}{\lam} &=& -{i\pi \over 2} (1-B)\lam 
=-{i\pi \over 2}(\lam -\lam^\dagger) \period \label{comRlam}
\end{eqnarray}

If we {\it do not} impose any constraints on $\lam$, then it 
 is quite easy to construct such an operator. Since 
 $\lam_A$ are all independent, we can introduce their conjugates 
$\omega_A$, carrying dimension 1,  which satisfy the simple OPE of 
the form\footnote{We follow the convention of 
Friedan-Martinec-Shenker\cite{FMS86} so that the sign is opposite to that of 
Berkovits.}
\begin{eqnarray}
\lam_A(z) \omega_B(w) &=& {\delta_{AB} \over z-w} \period \label{olope}
\end{eqnarray}
Then, $R$ can be defined as 
\begin{eqnarray}
R &=& {i\pi \over 2} \int [dz] :\lam_A(z) (1-B)_{AB} \omega_B(z):
\comma  \label{defR}
\end{eqnarray}
and (\ref{comRlam}) and hence (\ref{lamdag}) are realized. 

On the other hand, if one imposes the PS constraints, 
 construction  becomes  extremely difficult, if not impossible. 
Since $\lam_\Itil =(1/8)\lam_+^{-1} \ep_{IJKLM}\lam_{JK}\lam_{LM}$
 are now dependent composite fields, genuine conjugates to $\lam_A$ 
 do not exist. The closest analogue of $\omega_A$ is the field 
 introduced in \cite{Berk0001}, 
 denoted here by  $\tilde{\omega}_A$,
 which satisfies the OPE
\begin{eqnarray}
\lam_A(z) \tilde{\omega}_B(w) &=& {(1-K)_{AB} \over z-w} \comma \label{omlam}
\end{eqnarray}
with  $K_{AB}$  a projector needed for consistency with PS constrains. 
 Although $\tilde{\omega}$ and $K$ have a number of nice properties  and are extremely useful in the pure spinor formalism, 
one cannot simply substitute $\tilde{\omega}_B$ in place of $\omega_B$
 in (\ref{defR}). 
In fact the problem appears to be rather serious.
 Consider for example 
 the equation (\ref{comRlam}) for the component $\lam_\itil$. Since 
 $(\lam_\itil)^\dagger =\lam_{0i}$, we must have 
\begin{eqnarray}
\com{R}{\lam_\itil} &=& -{i\pi \over 2} (\lam_\itil -\lam_{0i}) \period 
\label{comRlami}
\end{eqnarray}
But since $\lam_\itil$ is already cubic in the basic fields, 
it is practically impossible  from the commutator 
to produce the term $\lam_{0i}$ on the RHS, 
 which is linear (or 
 quadratic in some parametrization \cite{Berk0001}) in
 the independent fields. We have not been able to construct $R$ which 
 effects the proper transformation (\ref{comRlam}) for all 
 the components consistently. 

This strongly indicates that in order to define an appropriate
 inner product with which one can properly implement the hermiticity property 
 and compute the amplitudes from the first principle,
 it appears to be imperative to remove the pure spinor 
 constraints. In the remainder of this paper, we will present a rather 
 elegant way of achieving this. 
\section{A New Closed First Class Algebra}
Besides the ones mentioned in the introduction, 
the analysis of the previous section added another reason to try to 
remove the PS constraints. 
In this and the subsequent sections, we shall show that it is indeed 
possible to achieve this by constructing a BRST-like charge, to be 
 called $\Qhat$, which is nilpotent without the PS constraints
 and  who's cohomology is identical to the constrained cohomology of $Q$. 
We begin, in this section,   by demonstrating
 that out of the system of second class 
 constraints formed by the basic operators $d_\al, \Pi^\mu$ and $\del
\theta^\al$ we can rather naturally construct, by using $U(5)$ 
 formalism,  a new closed first class algebra.
\subsection{$U(5)$ decompositions}
{}For this purpose, 
 we need to develop some tools to facilitate the manipulations in $U(5)$
 basis. 
Let $u^\mu$ be an $SO(9,1)$ vector. In the following, we will write 
 $u^0=iu^{10}$ and use $SO(10)$ notation. Hence the upper and the 
lower indices will 
 at times  not be distinguished.  Now $u^\mu$ can be decomposed 
 into two sets of 5-vectors $u^+_I$ and $u^-_I\comma (I=1\sim 5)$,
 which are $5$ and $\bar{5}$  of $U(5)$ respectively. This is 
done by introducing the \lq\lq intertwiners" $e^{\pm\mu}_I$ as
\begin{eqnarray}
u^{\pm}_I &=& e^{\pm\mu}_I u_\mu \comma 
\end{eqnarray}
where 
\begin{eqnarray}
e^{\pm\mu}_I &\equiv & \half (\delta_{\mu, 2I-1}\pm i\delta_{\mu,2I}) \period
\end{eqnarray}
It is easy to see that $e^{\pm \mu}_I$ enjoy the following basic 
properties:
\begin{eqnarray}
&& e^{\pm \mu}_I e^{\pm \mu}_J = 0 \comma \qquad 
  e^{\pm \mu}_I e^{\mp \mu}_J = \half \delta_{IJ} \comma \\
&&  e^{+\mu}_I e^{-\nu}_I +e^{-\mu}_I e^{+\nu}_I
 = \half \delta^{\mu\nu} \period
\end{eqnarray}
Using the last of these properties, $u^\mu$ can be expanded as
\begin{eqnarray}
u^\mu &=& 2\left(e^{+\mu}_I \um_I +e^{-\mu}_I \up_I \right) \period
\end{eqnarray}
In particular, the $\Ga$-matrices can be decomposed as 
\begin{eqnarray}
\Ga^\mu &=& 2\left(e^{+\mu}_I \Ga^-_I +e^{-\mu}_I \Ga^+_I \right) \comma 
\end{eqnarray}
where $\Ga^\pm_I$ are nothing but the fermionic oscillators 
 $b_I$ and $b_I^\dagger$:
\begin{eqnarray}
\Ga^+_I &=& e^{+\mu}_I \Ga^\mu = b_I \comma \qquad 
\Ga^-_I = e^{-\mu}_I \Ga^\mu = b_I^\dagger \period \label{GaI}
\end{eqnarray}

Next, we wish to  express  spinor bilinears in $U(5)$ basis.  Following the 
 method described in Appendix A, one can easily work out the formulas
 such as 
\begin{eqnarray}
e^{+\mu}_I (\lam \ga^\mu \chi) &=& -(\lam_\Jtil \chi_{IJ} + \lam_{IJ} 
\chi_\Jtil ) \comma \label{epgamu} \\
e^{-\mu}_I (\lam \ga^\mu \chi) &=& -(\lam_+\chi_\Itil+\lam_\Itil \chi_+)
 + {1\over 4} \ep_{IJKLM} \lam_{JK} \chi_{LM} \comma \label{emgamu}
\end{eqnarray}
where $\lam$ and $\chi$ are both chiral. 

Now we come to a simple but important observation.
 Let $\lam^\al$ be a bosonic chiral spinor
 and define 
\begin{eqnarray}
\Lam^\mu &\equiv & \lam \ga^\mu \lam \period \label{defLam}
\end{eqnarray}
10 conditions $\Lam^\mu=0$ make $\lam^\al$ to be a  pure spinor, 
 which actually has 11 independent components. This means that half 
 of the conditions must be redundant. To see this more explicitly, 
decompose $\Lam^\mu$ according to the formula (\ref{epgamu}) and 
 (\ref{emgamu}). We get
\begin{eqnarray}
\Lam^+_I &=& -2 \lam_{IJ} \lam_\Jtil \comma \\
\Lam^-_I &=& -2  \Phi_I \comma 
\end{eqnarray}
where
\begin{eqnarray}
\Phi_I &\equiv & \lam_+ 
\lam_\Itil -{1\over 8}  E_I \comma \label{defPhiI}\\
E_I &\equiv & \ep_{IJKLM}\lam_{JK}\lam_{LM} \period \label{defEI}
\end{eqnarray}
Evidently  $\Lam^-_I$ is directly proportional to the genuinely 
independent pure  spinor constraint  $\Phi_I$ but  $\Lam^+_I$ is not. 
However, by using (\ref{defPhiI}) we can rewrite it as\footnote{We assume $\lam_+\ne 0$, as in the original Berkovits formalism.}
\begin{eqnarray}
\Lam^+_I &=& -2\lampinv 
\lam_{IJ} \Phi_J -{1\over 4} \lam_+^{-1} \lam_{IJ} E_J\period
\end{eqnarray}
Now by a simple yet slightly non-trivial identity,  
 the second term on the RHS vanishes and we find that $\Lam^+_I$ is also 
a linear comibination of $\Phi_I$.  Summarizing,  we find that $\Lam^\mu$ can 
be decomposed naturally as 
\begin{eqnarray}
\Lam^\mu &=& \calN^\mu_I \Phi_I \comma \label{decompLam}
\end{eqnarray}
where 
\begin{eqnarray}
\calN^\mu_I &\equiv & -4(  e^{+\mu}_I -\lampinv \lam_{IJ}e^{-\mu}_J ) 
\label{defcalN}
\end{eqnarray}
are a set of 5 vectors. Moreover it is easy to check that they form a
 system of 5 {\it independent mutually orthogonal null vectors}, 
 consistent with the fact that $\Lam^\mu$ itself is null due to 
 the well-known Fierz identity
\begin{eqnarray}
(\ga^\mu)_{\al\be} (\ga_\mu)_{\ga\delta} + \mbox{cyclic in $(\al,\be,\ga)$} 
 =0\period \label{Fierz}
\end{eqnarray}

In fact, there exists another natural set of null vectors defined by 
\begin{eqnarray}
\calNbar^\mu_I &\equiv & -\half e_I^{-\mu} \period \label{defcalNbar}
\end{eqnarray}
$\calNbar^\mu_I$  are complimentary  to $\calN^\mu_I$ and they together
 satisfy 
the following orthonormality and completeness relations:
\begin{eqnarray}
\calN^\mu_I \calN^\mu_J &=& 0 \comma \qquad 
\calNbar^\mu_I \calNbar^\mu_J = 0\comma  \label{nuleq} \\
\calN^\mu_I \calNbar^\mu_J &=&  \delta_{IJ}\comma  \label{nnbar}\\
\calN^\mu_I \calNbar^\nu_I + \calN^\nu_I \calNbar^\mu_I 
 &=&  \delta^{\mu\nu} \period \label{nnbarsym}
\end{eqnarray}
These null vectors will play important roles\footnote{This is suspected to 
 be deeply related to the fact that pure spinors originally arose 
 in the description of null-planes\cite{Cartan}. It would be interesting
 to uncover the geometrical significance of our formalism.}. 
\subsection{New first class algebra}
Let us recall the set of OPE's among the basic superinvariant 
operators of dimension 1:
\begin{eqnarray}
d_\al(z) d_\be(w) &=& {2i\ga^\mu_{\al\be} \Pi_\mu (w)\over z-w}
\comma  \label{dd}\\
d_\al(z) \Pi^\mu(w) &=& {-2i(\ga^\mu \del\theta)_\al(w) \over z-w} \comma 
\label{dPi}\\
\Pi^\mu(z) \Pi^\nu(w) &=& -{\eta^{\mu\nu} \over (z-w)^2}
\comma  \label{PiPi}\\
d_\al(z) \del \theta^\be(w) &=& {\delta^\be_\al \over (z-w)^2} 
\period\label{ddelth}
\end{eqnarray}
The fact that the unit operators appear on the RHS of the last two OPE's 
 signify that they are of second class. With the help of the bosonic 
 chiral spinor variable $\lam^\al(z)$ and the decomposition (\ref{decompLam})
 involving null vectors, we can now turn this system into a first class 
 algebra without imposing any constraints on $\lam^\al$. 

First consider the OPE of the BRST current $j(z) =\lam^\al(z) d_\al(z)$
 with itself. Using (\ref{dd}) we immediately get
\begin{eqnarray}
j(z) j(w) &=& {2i\Lam^\mu \Pi_\mu(w) \over z-w} \period
\end{eqnarray}
Using the decomposition (\ref{decompLam}) and introducing a new operator
\begin{eqnarray}
\calP_I &\equiv & \calN^\mu_I \Pi_\mu \comma 
\end{eqnarray}
this can be written as
\begin{eqnarray}
j(z) j(w) &=& {2i\calP_I \Phi_I(w)
 \over z-w} \period
\end{eqnarray}
Now in contrast to the origial $\Pi^\mu$,  the operator $\calP_I$ has a 
first class OPE with itself due to the contraction with the null 
vector field $\calN^\mu_I$. Indeed, we have
\begin{eqnarray}
\calP_I(z) \calP_J(w)
 &=& \calN^\mu_I(z) \calN^\nu_J(w) \Pi_\mu(z)\Pi_\nu(w) \nn\\
&=& -{\calN^\mu_I(z) \calN^\mu_J(w) \over (z-w)^2} = {\calS_{IJ}(w) 
 \over z-w}\comma 
\end{eqnarray}
where the null nature of $\calN^\mu_I$ is crucial for the disappearance 
 of the double pole and 
\begin{eqnarray}
\calS_{IJ} &\equiv & -(\del \calN^\mu_I )\calN^\mu_J \period
\end{eqnarray}
Note that $S_{IJ}$ is properly antisymmetric again due to the null property
 of $\calN^\mu_I$. 

Consider next the OPE $j(z) \calP(w)$. We get
\begin{eqnarray}
j(z) \calP_I(w) &=& \lam^\al \calN^\mu_I d_\al(z) \Pi_\mu(w) 
= {-2i\calN^\mu_I (\lam \ga_\mu \del\theta) \over z-w}
\end{eqnarray}
Now we note the following useful representation of $\calN^\mu_I$:
\begin{eqnarray}
\calN^\mu_I &=& -2\lampinv (\ga^\mu \lam)_I \label{calNmu}
\end{eqnarray}
where $(\ga^\mu \lam)_I \equiv \bra{I} \Ga^\mu \ket{\lam} $. Then, 
 using the Fierz identity (\ref{Fierz}) one easily finds 
 $\calN^\mu_I (\lam \ga_\mu \del\theta) = \lampinv \Lam^\mu (\ga_\mu 
\del\theta)_I =\lampinv \calN^\mu_J  (\ga_\mu \del\theta)_I \Phi_J $. 
Further, it is not difficult to show that $ \calN^\mu_J 
 (\ga_\mu \del\theta)_I$ is actually anti-symmetric in $(I,J)$. 
Hence,  we get 
\begin{eqnarray}
j(z) \calP_I(w) &=& {\calR_{IJ} 
\Phi_J(w) \over z-w} \comma  \label{jcalP}
\end{eqnarray}
where 
\begin{eqnarray}
\calR_{IJ} &=& 2i\lampinv \calN^\mu_I (\ga_\mu \del\theta)_J \period
\label{defcalR}
\end{eqnarray}

In a similar manner, the rest of OPE's 
 between all the fields $j, \calP_I, \calR_{IJ}$
 and $S_{IJ}$ can be computed easily and we find that altogether
 they form the following set of Jacobi-consistent closed first class algebra:
\begin{eqnarray}
j(z) j(w) &=& {2i\calP_I \Phi_I \over z-w}\comma  \label{jj}\\
j(z) \calP_I(w) &=& {\calR_{IJ}\Phi_J \over z-w} \comma \label{jP}\\
j(z) \calR_{IJ} (w) &=& {-i\calS_{IJ} \over z-w} \comma \label{jR}\\
\calP_I(z) \calP_J(w) &=& {\calS_{IJ} \over z-w} \comma \label{PP}\\
\mbox{all the rest} &=& \mbox{non-singular} \period \nn
\end{eqnarray}

So with the help of unconstrained $\lam^\al$, we have been able to 
 turn the system with second class constraints into one which is 
 of purely first class. 

Let us make some remarks. First, an algebra similar in spirit to the above 
 appeared in the analysis of BRST cohomology in $SO(8)$ framework
\cite{Berk0006}. The system itself was much simpler, consisting 
 of two operators called $G^a$ and $T$, but 
as the $SO(8)$ parametrization of the PS constraints is highly redundant, 
 these operators were quite complicated containing inifinite number of 
 ghosts. In our case, 
 as we have captured the content of the PS constraints without 
 redundancy, the operators are  simple with no ghosts required. 
Second, the method we have developed can be applied to more general 
 systems of second class algebra. Such applications may be useful 
 in many other contexts and will be described in a separate publication 
\cite{AK3}. 
\section{Homological Perturbation, Nilpotent BRST-like \\
 Charge and its 
Cohomology}
\subsection{Batalin-Vilkovisky procedure}
In the previous section, we have obtained in a rather natural manner
 a new closed first class algebra. An immediate thought which comes to one's
 mind is to apply the usual BRST formalism of Batalin and Vilkovisky (BV)
\cite{BatVilk}
 to construct a nilpotent BRST-like charge associated with this algebra. 

This can indeed be done, albeit with a slight peculiarity. The peculiarity 
 is that the operators which form the algebra are not the usual 
 constraints which generate  the underlying classical 
gauge symmetry\footnote{In fact this is the reason why we have been careful to 
 avoid calling it a \lq\lq constraint" algebra.}.  
In particular, they include the current $j(z)$, which
 if $\lam^\al$ were a pure spinor is interpreted as the nilpotent BRST-like
 current of the system. Nevertheless, BV procedure can be applied as 
 a formal algorithmic device to construct a fermionic charge, to be 
 called $Q'$, which is nilpotent without pure spinor constraints. 
As usual we first introduce the ghost-antighost pair for each operator forming 
 the algebra in the following way\footnote{Notation for the anti-ghost
 $b_I$ coincides  with the fermionic oscillator used to in $U(5)$ formalism, 
but there should be no confusion.} :
\begin{eqnarray}
j &:& \quad (\ga, \be)\comma  \\
\calP_I &:& \quad (c_I, b_I) \comma \\
\calR_{IJ} &:& \quad (\ga_{IJ}, \be_{IJ})\comma  \\
\calS_{IJ} &:& \quad (c_{IJ}, b_{IJ}) \period
\end{eqnarray}
Here 
$(\ga,\be)$ ghosts are bosonic and $(c,b)$ pairs are fermionic, 
both carrying the conformal weights $(0,1)$, with the 
 OPE of the form 
\begin{eqnarray}
b_I(z) c_J(w) &=& {\delta_{IJ} \over z-w} \comma \\
\beta_{IJ}(z) \ga_{KL} &=& -{\delta^{IJ}_{KL} \over z-w} \qquad etc.
\end{eqnarray}
As for the central charge counting, $(\ga_{IJ}, \be_{IJ})$ and 
$(c_{IJ}, b_{IJ})$ compensate, while $(c_I, b_I)$ precisely kill 
 the contribution from the extra 5 components in $\lam^\al$ 
(and their conjugates) now alive in the absence of the PS constraints. 
One might worry that 
the contribution from $(\ga,\be)$ ghosts remains uncancelled. 
However, because $j(z)$ does not appear on the RHS of the OPE's, $\be$ will be
 absent in $Q'$ and hence $\ga$ can be simply  set to 1. Therefore
 the total central charge still vanishes and we have a viable free conformal 
 field theory. 

Following the standard BV prescription, the BRST charge $Q'$ 
 is now constructed as $Q'=\int [dz] j'(z)$,  where
 the BRST current is given by 
\begin{eqnarray}
j' &=& j + c_I \calP_I + \half \ga_{IJ} \calR_{IJ} 
+\half c_{IJ} \calS_{IJ} \nn\\
&& \quad -ib_I \Phi_I +\beta_{IJ} \Phi_J c_I +{i\over 2} b_{IJ} \ga_{IJ}
+\half b_{IJ}c_Ic_J \period \label{jprime}
\end{eqnarray}
It is easy to check that $Q'$ is indeed nilpotent. 

The crucial question of course is whether the cohomology of $Q'$ is 
 isomorphic to the constrained cohomology of $Q$. The answer turned out
 to be yes, but
 we will not give the details of the proof here. 
The reason is that, as we shall 
 shortly describe, there exists an alternative scheme of producing a nilpotent 
 BRST-like charge which is much more natural and profound in the present 
 context than the BV procedure. 
It is known under the name of {\it homological perturbation} 
\cite{HenTb}. 

Nevertheless, it may be instructive to mention briefly how we were led to 
 the use of the homological perturbation scheme. In the effort to demonstrate
 the equivalence of the aforementioned cohomologies, we first proved 
 that any expressions involving $(\ga_{IJ}, \be_{IJ})$ and 
$(c_{IJ}, b_{IJ})$ ghosts are actually $Q'$-exact and hence can be dropped. 
This is quite natural as these ghosts form  quartets. 
This means that effectively $Q'$ can be reduced to a much 
 simpler $\tilde{Q}$ given by 
\begin{eqnarray}
\tilde{Q} &=& Q + \delta + d_1\comma  \label{Qtil}
\end{eqnarray}
where 
\begin{eqnarray}
\delta &\equiv & \int [dz] \left( -i\Phi_I  b_I \right) 
\comma 
\label{delta} \\
d_1 &\equiv & \int [dz] c_I \calP_I \period \label{done}
\end{eqnarray}
The notations here are designed to imply that $\delta, Q$ and $d_1$ 
 can be regarded as $-1, 0$ and 1 form operators respectively. 
As we have truncated the full $Q'$, $\acom{\tilde{Q}}{\tilde{Q}} $ 
 no longer vanishes but equals $\int [dz] (-2c_IR_{IJ}\Phi_J 
 + c_Ic_JS_{IJ} )$. However, this quantity commutes with 
 any expressions composed of $\lam^\al, x^\mu, \theta^\al, c_I$ and 
their worldsheet derivatives,
 which are the building blocks of the unintegrated 
vertex operators. Moreover
 the pieces neglected in $Q'$ have no singularities with such 
 operators. Thus the cohomology analysis can indeed be carried out with 
 $\tilde{Q}$ and in this way we succeeded in  producing a proof. 

At this point an alert reader may have noticed that the form of $\Qtil$ 
is precisely the beginning of the homological perturbation scheme, 
 which allows one to upgrade $\Qtil$ into a genuinely nilpotent 
 operator $\Qhat$ under appropriate conditions.
 Moreover the general theory guarantees that 
 the cohomology of $\Qhat$ is equivalent to the cohomology of $Q$ 
 with the constraint $\delta =0$, which in our case is nothing but 
 the imposition of PS constraints. In other words, while BV procedure 
 provides  one way of nilpotent completion of $Q$, 
the homological perturbation,  to be described below,  gives another, 
 which in our case is more natural and powerful. As we shall see, 
 this in general leads to a completion different from the one obtained 
 by the BV method. 
\subsection{Homological perturbation and  nilpotent charge $\Qhat$}
Now let us explain the homological perturbation scheme, as applied 
 to our system. Since the general theory is lucidly described in \cite{HenTb}, 
we shall limit the exposition to the extent necessary for our purposes, 
which includes the application to the construction of physical vertex 
operators described in Sec.6. 

As was already mentioned, we will use the terminology of the 
differential form, where the degree of the form is defined as 
 the total $(c_I,b_I)$ ghost number, with ${\rm gh \#}(c_I) =1, 
{\rm gh \#}( b_I) =-1$.  Also, in the following,
 a product $AB$ will always 
signify the operator product in the sense of conformal field theory. 
In particular, when $A$ or $B$ is an integrated operator, $AB$
 equals the graded commutator $\gcom{A}{B}$. In this notation, 
 the graded Jacobi identity reads $ABC \equiv A(BC) = (AB)C \pm B(AC)$. 

Now in the homological perturbation theory, the 
 operator $\delta$, defined in our case in (\ref{delta}), of degree $-1$ 
will play the key role. 
It must satisfy the following basic properties:
\begin{eqnarray}
&& (i)\qquad \delta^2 = 0 \comma \label{delsqzero} \\
&&(ii)\qquad \delta A = 0 \quad \Rightarrow \quad A =\delta B \comma 
\label{delcohom}
\end{eqnarray}
where $A$ and $B$ are, respectively, an $n$-form and an $(n+1)$-form 
 with $n\ge 1$. The property $(i)$, which in our case is obvious from the 
 definition of $\delta$, means that one can consider the
 {\it $\delta$-homology}
 defined by $H_n(\delta) = {\rm Ker}\ \delta/ {\rm Im}\ \delta$, where 
 $n$ is the degree of the space on which $\delta$ acts. Then 
 the property $(ii)$ simply states that $H_n(\delta) =0$  for 
$n \ge 1$, namely the homology is trivial above degree 1. 
This property is absolutely crucial for the whole scheme to work and hence
 we shall give the proof for our  $\delta$. 

For the simple case of operators 
 consisting of $x^\mu, \theta^\al, \lam^\al$ and their worldsheet
 derivatives together with  {\it $c_I$'s without any $\del$},
 the proof is straightforward 
  since in such a situation the action of $\delta$ is simply to 
replace one $c_I$ by $\Phi_I$. 
Let $A=c_{I_1}c_{I_2}\cdots 
c_{I_n} A_{I_1I_2\ldots I_n}$ be an $n$-form. Acting $\delta$, we get
$\delta A =-in c_{I_2}c_{I_3}\cdots c_{I_n} \Phi_{I_1} 
A_{I_1 I_2I_3 \ldots I_n}$. 
Because we have captured the PS constraints without redundancy 
{\it $\Phi_I$'s are algebraically independent}. Therefore for $\delta A_n$ 
  to vanish we must have
$A_{I_1 I_2I_3 \ldots I_n} =\Phi_J \tilde{A}_{J I_1I_2\ldots I_n } $, where 
 $\tilde{A}$ is totally antisymmetric. But then 
$A = \Phi_J c_{I_1} c_{I_2} \cdots c_{I_n} \tilde{A}_{JI_1I_2 \ldots I_n}$
 can be written as $\delta B$, where $B=(i/(n+1)) c_Jc_{I_1} c_{I_2} \cdots c_{I_n} \tilde{A}_{JI_1I_2 \ldots I_n} $. 

The proof for the general case is considerably more involved. 
The reason is that, in addition to the aformentioned building blocks, 
we may have $b_I, \omega_\al$, worldsheet derivatives thereof and 
$\del^m c_I$'s. The OPE of $\delta$ with such operators can produce 
higher order poles as well as structures other than $\Phi_I$'s. 
 Nevertheless by appropriate use of mathematical inductions the proof can be
 produced, which is given in Appendix B. 

Besides $\delta$, another operator of prime importance is, of course,
 the operator $Q$. Clearly $Q$ satisfies 
\begin{eqnarray}
 \delta Q =0 \period \label{deltaQ}
\end{eqnarray}
In other words, $Q$ anticommutes with $\delta$. 
Another important property of $Q$ is that while 
 $Q^2$ does not vanish  it is nevertheless
 $\delta$-exact. Indeed, 
\begin{eqnarray}
Q^2 = \int[dz] 2i \calP_I \Phi_I = -2\delta d_1 \comma \label{Qsq}
\end{eqnarray}
where $d_1$ is given in (\ref{done}). In homological perturbation 
theory, $Q$ satisfying (\ref{deltaQ}) and (\ref{Qsq}) 
 is said to be {\it a differential modulo $\delta$}. 

Now with this setting  one can 
 construct a nilpotent operator $\Qhat$ in the manner
\begin{eqnarray}
\Qhat &=& \delta + Q + d_1 + d_2 + \cdots \comma  \label{Qhat}
\end{eqnarray}
where $d_n$ is an operator of degree $n$. Moreover, the main theorem 
 of homological perturbation theory states \cite{HenTb} 
that the cohomology of $\Qhat$
 coincides with the $\delta$-constrained cohomology of $Q$. 

To find $d_n$, we first write down the requirement of nilpotency of $\Qhat$ 
more explicitly:
\begin{eqnarray}
0 &=& \Qhat^2 = Q^2 + 2\sum_{n\ge 1} \delta d_n + 2\sum_{n\ge 1} Qd_n 
 + \sum_{k,l\ge 1} d_kd_l \period \label{nilpQhat}
\end{eqnarray}
Rather than repeating the general recursive procedure given
 in \cite{HenTb}, let us
 see how  this equation determines $d_n$ explicitly in our context. 
At overall degree 0, the nilpotency is fulfilled due to (\ref{Qsq}). 
At degree 1, the condition becomes 
\begin{eqnarray}
0 &=& \delta d_2 + Qd_1 \period \label{leveloneeq}
\end{eqnarray}
Now apply $\delta$ on the second term $Qd_1$. Using (\ref{deltaQ}), 
 a graded  Jacobi identity and (\ref{Qsq}), 
we get $\delta (Qd_1) = -Q(\delta d_1) =\half Q^3=0$. {}From 
 (\ref{delcohom}) this means that $Qd_1$ must be of the form 
$Qd_1 =-\delta X_2$, where  $X_2$ is a 2-form. Then, (\ref{leveloneeq})
 becomes  $0 =\delta (d_2-X_2)$ and this is solved for $d_2$ as 
$d_2 =X_2 + \delta Y_3$ 
with an arbitrary 3-form $Y_3$. Note that all the manipulations are 
 independent of the details of the operators. The explicit content 
 of the operators does matter when we actually determine the form of $X_2$ 
 etc. In our case, we have
\begin{eqnarray}
Qd_1 &=& -\int [dz] c_I R_{IJ} \Phi_J = -\delta X_2 \nn\\
X_2 &=& -{i\over 2}\int [dz]  c_Ic_J R_{IJ} \period
\end{eqnarray}

Let us go one more step to the degree 2 analysis. The nilpotency condition
 now reads
\begin{eqnarray}
0 &=& 2\delta d_3 + 2Qd_2 + d_1^2 \period
\end{eqnarray}
Explicit calculation immediately gives $2QX_2=-c_Ic_J \calS_{IJ}$ 
 and  $d_1^2 = c_Ic_J \calS_{IJ}$. Therefore the main part of 
$Qd_2+d_1^2$ vanishes and we are left with 
$\delta (d_3-QY_3)=0$. But since $Y_3$ is arbitrary, we may set it to 
 zero and this gives $d_3=0$ as a viable solution. In this way, 
 for the system at hand the perturbation terminates at this stage. 
Summarizing, we now have constructed a nilpotent operator $\Qhat$ 
 in the form 
\begin{eqnarray}
\Qhat &=& \delta + Q + d_1 + d_2 \comma \label{Qhattot}
\end{eqnarray}
where $\delta, Q, d_1$ were already given before and $d_2$ can be taken to be
\begin{eqnarray}
d_2 = -{i\over 2} \int [dz] c_Ic_J R_{IJ} \period \label{dtwo}
\end{eqnarray}

At this point, let us introduce  \lq\lq semi-covariant" notations 
 for the ghosts. Define $c^\mu$ and $b^\mu$ by
\begin{eqnarray}
c^\mu &\equiv & c_I\calN^\mu_I \comma \qquad
b^\mu \equiv  b_I\calNbar^\mu_I \period \label{cbmu}
\end{eqnarray}
Using the orthonormality relations for $\{\calN^\mu_I, \calNbar^\mu_I\}$
 these relations can be inverted as 
\begin{eqnarray}
c_I &=& c^\mu \calNbar^\mu_I\comma \qquad b_I = b^\mu \calN^\mu_I \period
\label{cbI}
\end{eqnarray}
They satisfy the OPE
\begin{eqnarray}
b^\mu(z) c^\nu(w) &=& {\calNbar^\mu_I\calN^\nu_I(w) \over z-w}
 = {\eta^{\mu\nu}-\calNbar^\nu_I \calN^\mu_I(w) \over z-w} \period
\end{eqnarray}
We emphasize that $b^\mu$ and  $c^\mu$ are not genuine vector fields 
and this fact is reflected in this OPE. Nevertheless, for computational 
 purposes this semi-covariant notation will be useful. In this notation,
the operators $\delta, d_1$ and $d_2$ take simple forms:
\begin{eqnarray}
\delta &=& -\int [dz] ib^\mu \Lam_\mu \comma \quad 
d_1 = \int [dz] c^\mu \Pi_\mu\comma \\
d_2 &=& \int[dz] \lampinv c^\mu c^\nu \calNbar^\nu_I(\ga_\mu \del\theta)_I
\end{eqnarray}

Some remarks are in order. 
(1)\ The scheme of homological perturbation can be thought of as 
a device to implement {\it the constraints $\Phi_I=0$  in  $\lam$-space}
 by enlarging the space to include $(c_I, b_I)$ ghosts, in contrast 
 to the usual BV procedure which is normally used to implement 
 the original gauge constraints. 
(2)\ The ambiguity such as the choice of $Y_3$
 above can be utilized to obtain different form of $\Qhat$ if one 
 desires. 
(3)\ With our simplest choice of fixing this ambiguity, only 
 a minimum number of ghosts $(c_I,b_I)$ are needed. 

\subsection{Equivalence of cohomologies}
Although the equivalence of the cohomology of $\Qhat$ to the 
 constrained cohomology of $Q$ is guaranteed by the general 
 theory of homological perturbation \cite{HenTb}, we shall give a sketch 
 of the proof, as the similar logic will be needed in the construction of 
 the vertex operators. 

In the proof, the nilpotency of $\Qhat$ will play an important role. 
By sorting out the equation $\Qhat^2=0$ (\ref{nilpQhat})
according  to the degree, we have, in our case,  the explicit relations
\begin{eqnarray}
\delta^2 &=&0 \comma \label{delsq}\\
\delta Q &=& 0 \comma \label{delQ}\\
Q^2 + 2\delta d_1 &=& 0 \comma \label{Qsqp}\\
Qd_1 + \delta d_2 &=& 0 \comma \label{Qdonep}\\
d_1^2 + 2Qd_2 &=& 0 \comma \label{donesqp}\\
d_1d_2 &=& 0 \comma \label{donedtwo}\\
d_2^2 &=& 0 \period \label{dtwosq}
\end{eqnarray}
It is also useful to write down the action of $\Qhat$ on a general 
 operator $X$, which can be expanded with respect to the degree\footnote{
Since $c_I$ has 5 components, the highest possible degree is 5.}
 as $X=\sum_{n=0}^5 X_n$: 
\begin{eqnarray}
\Qhat X &=& \sum_n (QX_n + d_1X_{n-1} + d_2X_{n-2} + \delta X_{n+1} )
 \period \label{QhatX}
\end{eqnarray}

Now we begin with the proof of $Q$-closed $\Rightarrow$ $\Qhat$-closed. 
Let $U_0$ be an operator which is $Q$-closed up to the PS constraint, \ie 
$Q U_0 = U_I \Phi_I$ for some $U_I$. Since the RHS can be rewritten as 
$-\delta U_1$, where  $U_1 = -ic_I U_I$, we have 
$QU_0 +\delta U_1=0$, which, according to (\ref{QhatX}) 
is the $\Qhat$ closedness relation at degree 0. Of course $U_1$ is 
 determined only up to some $\delta X$, but such a freedom is easily seen 
 to correspond precisely to that of adding a $\Qhat$-exact form and 
 hence will be ignored hereafter. Suppose that we have constructed 
 $U_k$'s up to $k=n+1$, which satisfy the $\Qhat$-closedness relation 
\begin{eqnarray}
QU_n + d_1 U_{n-1} +d_2 U_{n-2} +\delta U_{n+1} =0 \period 
\label{Qhatcln}
\end{eqnarray}
Our task is to construct $U_{n+2}$ which satisfies the similar 
 equation at one degree higher. Consider the expression 
 $Y_{n+1} \equiv QU_{n+1}  + d_1 U_n +d_2U_{n-1}$. Act $\delta$ onto this and 
make use of the relation (\ref{delQ}). We get 
\begin{eqnarray}
 \delta Y_{n+1} 
 = -Q(\delta U_{n+1}) + \delta(d_1U_n) + \delta(d_2 U_{n-1}) 
\period
\end{eqnarray}
Applying the relation (\ref{Qhatcln}) to the first term on the RHS and 
 using Jacobi identities such as $\delta(d_1U_n) = (\delta d_1) U_n 
 -d_1(\delta U_n)$, $Q(QU_n) = \half Q^2 U_n$, etc., this can be
 rewritten into 
\begin{eqnarray}
\delta Y_{n+1} &=& -d_1(QU_{n-1} + d_1 U_{n-2} + \delta U_n)
 -d_2(QU_{n-2} + \delta U_{n-1}) \period 
\end{eqnarray}
Now substituting  the relations (\ref{Qhatcln}) at lower degrees, this becomes 
\begin{eqnarray}
\delta Y_{n+1} &=& d_1(d_2U_{n-3}) + d_2(d_1U_{n-3} + d_2U_{n-4})\comma 
\end{eqnarray}
and this is seen to  vanish by  (\ref{donedtwo}) and (\ref{dtwosq}). 
This means that  $Y_{n+1}$ must be of the structure $Y_{n+1} = -\delta 
 U_{n+2}$ and in this way we can construct $U_{n+2}$ that satisfies 
 the $\Qhat$-closedness condition at degree $n+2$. 
Hence $U =\sum_n U_n$ is $\Qhat$-closed  and the proof is complete. 

The logic for proving $Q$-exact $\Rightarrow$ $\Qhat$-exact is a slight 
 variation of the above. 
Suppose $U_0$ is $Q$-exact, \ie $U_0 = Q\Omega_0 + \delta \Omega_1$ for 
 some $\Omega_0$ and $\Omega_1$. Applying $Q$ to this relation, 
 one easily gets $QU_0 = -\delta (Q\Omega_1 +d_1\Omega_0)$. 
Comparing this with (\ref{Qhatcln}) which  an exact form, 
 necessarily being closed, must also satisfy, we can set 
$U_1 = Q\Omega_1 +d_1\Omega_0+
\delta \Omega_2$ with some $\Omega_2$. Referring to 
 (\ref{QhatX}), this in turn means that $U_0+U_1$ is $\Qhat$-exact up to 
 the degree in question. In an entirely similar manner, one can successively
 construct  $U_n$ in the form $U_n = Q\Omega_n + d_1\Omega_{n-1} 
 + d_2 \Omega_{n-2} + \delta \Omega_{n+1}$ and hence finds
$U = \sum_n U_n =\Qhat \Omega$, where $\Omega \equiv \sum_n \Omega_n$. 

The converses to these statements are much easier to prove. To show that 
 $\Qhat$-closed $\Rightarrow$ $Q$-closed, let $U=\sum_n U_n$ be 
 $\Qhat$-closed and consider the projection map $U\rightarrow U_0$. 
Then, $U_0$ is $Q$-closed since it satisfies $QU_0+\delta U_1=0$. Similarly,
 if we let $U=\sum_n U_n =\Qhat \Omega =\sum_n \Qhat \Omega_n$, then $U_0$ 
 is $Q$-exact as it satisfies $U_0=Q\Omega_0 + \delta \Omega_1$. 
This shows $\Qhat$-exact $\Rightarrow$ $Q$-exact, and the entire 
 proof of the equivalence of cohomologies is completed. 
\subsection{Action and energy-momentum tensor}
Now that we understand what fields are needed for our formalism,
let us summarize their properties and 
 write down the free action and the energy-momentum 
 tensor. Besides the string coordinate $x^\mu$, we have 
 the conjugate pairs $(\theta^\al, p_\al)$, $(\lam^\al, \omega_\al)$
 and $(c_I, b_I)$ all carrying  conformal dimensions $(0,1)$. 
The  Euclidean action is given by 
\begin{eqnarray}
S &=& {1\over \pi} \int d^2z \left(\half \del x_\mu \bar{\del}x^\mu
 +p_\al \bar{\del} \theta^\al + \omega_\al \bar{\del} \lam^\al
 +b^\mu \bar{\del} c_\mu   \right) \comma \label{action} 
\end{eqnarray}
where we used the semi-covariant notations for the ghosts. The
 associated the energy-momentum tensor is 
\begin{eqnarray}
T &=& -\half \del x^\mu \del x_\mu -p_\al \del\theta^\al 
 -\omega_\al \del \lam^\al -b^\mu \del c_\mu \nn\\
&=& -\half \Pi^\mu \Pi_\mu -d_\al\del\theta^\al
 -\omega_\al \del \lam^\al -b^\mu \del c_\mu
 \period \label{emtensor}
\end{eqnarray}
It is clear that the system forms a free CFT with vanishing central charge. 

Since we have so far introduced only the minimum number of fermionic 
 ghosts, they are not yet Lorentz invariant. 
A simple way to remedy 
 this would be to introduce another five pairs 
of ghosts $(\tilde{c}_I, \tilde{b}_I)$ to promote our semi-covariant 
 $c^\mu, b^\mu$ to genuine  Lorentz vectors, as in \cite{stonybrook}.
Of course one needs to add additional ghosts to cancel the extra 
central charge produced by this revision. 
 How this  should be done, without spoiling the cohomology, 
 to make the theory manifestly Lorentz covariant is under study 
 and will be discussed in a separate publication \cite{AK2}. 
\section{Vertex Operators for Massless Modes}
In order to compute the scattering amplitudes, one needs to construct 
 the physical vertex operators. In this work,  for simplicity 
we shall restrict ourselves to those for the super-Maxwell multiplet. 
Extension to the super-Yang-Mills case and to the closed string 
 massless modes should be straightforward. 
\subsection{Unintegrated vertex by homological perturbation}
Let us start with the construction of 
unitegrated vertex operator, to be denoted by 
 $U$. With pure spinor constraints, it has been shown \cite{Berk0001, 
Siegel86} that the 
 massless unintegrated vertex operator is given by 
\begin{eqnarray}
U_0 &=& \lam^\al A_\al(x,\theta) \comma \label{Uzero}
\end{eqnarray}
where $A_\al$ is the spinor superfield that satisfies 
 the on-shell condition 
\begin{eqnarray}
(\ga^{\mu_1\mu_2\ldots \mu_5})^{\al\be} D_\al A_\be =0 \period 
\label{onshellcond}
\end{eqnarray}
As is well-known \cite{onshellcond}, one can derive the following important 
 equations from this condition, to be frequently used in the 
subsequent analysis:
\begin{eqnarray}
&&(i)\quad D_\al A_\be +D_\be A_\al = -2i\ga^\mu_{\al\be} B_\mu 
 \comma \label{OSi} \\
&&(ii)\quad D_\al B_\mu -\del_\mu A_\al = -
2i(\ga_\mu)_{\al\be} W^\be  \comma  \label{OSii} \\
&&(iii)\quad D_\al W^\be =-{1\over 4} (\ga^{\mu\nu})^\be{}_\al F_{\mu\nu} 
 \comma  \label{OSiii} \\
&& (iv)\quad D_\al \Fmn = 2i((\ga_\mu\del_\nu W)_\al 
-(\ga_\nu\del_\mu W)_\al) \comma \label{OSiv} \\
&& (v)\quad \ga^\mu \del_\mu W^\al =0 \comma \label{OSv}\\
&& (vi)\quad \del^\mu \Fmn =0 \period \label{OSvi} 
\end{eqnarray}
{}From $(i) \sim (iii)$, one finds 
\begin{eqnarray}
B_\mu &=& {i\over 16} \ga_\mu^{\al\be} D_\al A_\be \comma \label{defB} \\
W^\al &=& {i\over 20} (\ga^\mu)^{\al\be} (D_\be B_\mu -\del_\mu A_\be)
\comma \label{defW} \\
F_{\mu\nu} &=& \del_\mu B_\nu -\del_\nu B_\mu \period \label{defF}
\end{eqnarray}

 In our formalism without 
 pure spinor constraints, $U_0$ must be extended to include additional pieces 
 contaning $c_I$  ghosts so that the total vertex operator $U$ 
 satisfies $\Qhat U=0$. The method of construction should already be 
 clear from the cohomology analysis presented in the previous section. 
Namely, we will expand the dimension 0 operator 
$U$ in powers of $c_I$ in the form 
 $U=\sum_{n=0}^5 U_n$, where $U_n$ contains products of $n$ $c_I$'s, 
 and impose $\Qhat U=0$ to fix the coefficient fields, up to 
 a $\Qhat$-exact form. In this analysis,  $d_2$ term in $\Qhat$ can be
 dropped since it produces no singularity in the OPE's 
with any other quantities involved. Thus, the master equation at 
  degree $n$ takes the form 
\begin{eqnarray}
QU_n + \delta U_{n+1} + d_1 U_{n-1} =0 \period \label{unintmastereq}
\end{eqnarray}
{}From the general cohomology analysis we know that a consistent
 solution exists at every degree. More concretely, this means 
that $QU_n+ d_1 U_{n-1}$ is guranteed to be $\delta$-exact so that 
 $U_{n+1}$ can be found by using the triviality of the $\delta$-homology.  

Let us begin at degree 0. The master equation in this case is  
\begin{eqnarray}
Q U_0 + \delta U_1 &=& 0 \period \label{deg0eq}
\end{eqnarray}
Using the equation (\ref{OSi})  we immediately get
\begin{eqnarray}
QU_0 &=& -i\Lam^\mu B_\mu  \period 
\end{eqnarray}
Note that  $QU_0$ vanishes with PS constraints $\Lam^\mu=0$, as it should. 
Then (\ref{deg0eq}) is easily solved to give 
\begin{eqnarray}
U_1 &=& -c^\mu B_\mu \period \label{Uone}
\end{eqnarray}

At degree 1, the basic equation takes the form 
\begin{eqnarray}
QU_1 + \delta U_2 + d_1 U_0 =0 \period \label{deg1eq}
\end{eqnarray}
A simple calculation using (\ref{OSii}) yields 
\begin{eqnarray}
QU_1 + d_1 U_0 
&=& -2i  c^\mu  (\lam \ga_\mu W) \comma \label{QUonep}
\end{eqnarray}

We recognize that the expression on the RHS of (\ref{QUonep})
 is identical to 
 the one we encountered in Sec.4, with $\del\theta^\al$ in place
of $W^\al$, and can be rewritten as 
\begin{eqnarray}
QU_1 + d_1 U_0 &=& -2i\lampinv c^\nu \Lam^\mu \calNbar^\nu_I (\ga_\mu W)_I 
 \period
\end{eqnarray}
This  is manifestly proportional to $\Lam^\mu$ and hence can be easily written 
  as $-\delta U_2$. In this way we obtain
\begin{eqnarray}
U_2 &=& -\lampinv c^\mu c^\nu \calNbar^\nu_I (\ga_\mu W)_I
  \period   \label{Utwo}
\end{eqnarray}

The construction at degree 2 is a little more involved. The equation 
 to solve is 
\begin{eqnarray}
QU_2 + \delta U_3 + d_1 U_1 =0  \comma \label{deg2eq}
\end{eqnarray}
and again to get $U_3$ we must compute $QU_2 + d_1 U_1$ and write it 
 as a $\delta$-exact form. For this calculation, we need Eq.(\ref{OSiii}), 
 the Fierz identity (\ref{Fierz}) 
 as well as an identity  $c^\mu (\ga_\mu \ga_{\nu\rho}\lam)_I 
 =\lampinv c^\sigma \calNbar^\sigma_J \Lam^\mu (\ga_{\mu\nu\rho})_{IJ} 
-\lam_+ (c_\nu \calN^\rho_I -c_\rho \calN^\nu_I)$, where  
$(\ga_{\mu\nu\rho})_{IJ}\equiv \bra{I} \Ga_{\mu\nu\rho} \ket{\Jtil} 
= 64 \ep_{IJPQR} \calNbar^\mu_P \calNbar^\nu_Q \calNbar^\rho_R$. 
In this way, we get 
\begin{eqnarray}
QU_2 + d_1U_1 &=& -{1\over 4} \lam_+^{-2} \Lam^\mu c^\sigma c^\tau 
\calNbar^\sigma_I \calNbar^\tau_J (\ga_{\mu\nu\rho})_{IJ} F^{\nu\rho} 
 \period \label{QUtwopp}
\end{eqnarray}
As this is again proportional to $\Lam^\mu$, 
 it is now a simple matter 
 to express the RHS of (\ref{QUtwopp}) in the form  $-\delta U_3$ 
 and we obtain 
\begin{eqnarray}
U_3 &=& {i\over 12} \lam_+^{-2} c^\mu c^\sigma c^\tau 
\calNbar^\sigma_I \calNbar^\tau_J (\ga_{\mu\nu\rho})_{IJ} F^{\nu\rho} 
\period \label{Uthree}
\end{eqnarray}

The master equation at degree 3, \ie $QU_3+d_1U_2+\delta U_4=0$, 
 can be analyzed in a similar manner. 
Omitting the details, we find 
\begin{eqnarray}
U_4 &=& {i\over 12}\lam_+^{-3}  c^{\mu_1}c^{\mu_2}c^{\mu_3}c^{\mu} 
\calNbar^{\mu_1}_{I_1} \calNbar^{\mu_2}_{I_2} \calNbar^{\mu_3}_{I_3}
(\ga_{\mu\nu\rho})_{I_1I_2} (\ga^\nu \del^\rho W)_{I_3}
\period \label{Ufour}
\end{eqnarray}

Now at degree 4, where we have to solve $QU_4 + d_1U_3 + \delta U_5=0$, 
 we found that $QU_4+dU_3$ vanishes identically using the 
on-shell condition as well as the Bianchi identity for $F_{\mu\nu}$. 
Thus we get the simple result $U_5=0$. Then the final equation 
 at degree 5, which serves as the consistency condition, 
 reduces to $d_1 U_4=0$ and it can be checked that
 this indeed holds. 

Summarizing, we have found that the unintegrated vertex 
operator $U$ for the super-Maxwell multiplet is given, up to 
 a $\Qhat$-exact form,  by 
\begin{eqnarray}
U &=& \sum_{n=0}^4 U_n \comma \\
U_0 &=& \lam^\al A_\al\comma  \\
U_1 &=& -c^\mu B_\mu \comma \\
U_2 &=& -\lampinv c^\mu c^\nu \calNbar^\nu_I (\ga_\mu W)_I \comma \\
U_3&=& {i\over 12} \lam_+^{-2} c^\mu c^\sigma c^\tau 
\calNbar^\sigma_I \calNbar^\tau_J (\ga_{\mu\nu\rho})_{IJ} F^{\nu\rho} \comma \\
 U_4 &=& {i\over 12}\lam_+^{-3}  c^{\mu_1}c^{\mu_2}c^{\mu_3}c^{\mu} 
\calNbar^{\mu_1}_{I_1} \calNbar^{\mu_2}_{I_2} \calNbar^{\mu_3}_{I_3}
(\ga_{\mu\nu\rho})_{I_1I_2} (\ga^\nu \del^\rho W)_{I_3}
\period 
\end{eqnarray}
Evidently, $U$ is not manifestly Lorentz invariant. However, since 
 the cohomology of $\Qhat$ is equivalent to that of $Q$ with 
 pure spinor constraints, and since the latter has been shown to 
 respect Lorentz covariance \cite{Berk01051}, it should be possible 
 to extend our formalism, with additional ghosts, 
to make  the Lorentz covariance manifest. 
\subsection{Relation to Virasoro operator and construction of 
integrated vertex}
Having obtained the unitegrated vertex, our next task is to
 construct the integrated vertex operator, to be denoted by 
 $\int[dz]V(z)$,  which is required for computation of 
$n$-point amplitudes with $n \ge 4$. $V(z)$ of dimension 1
 is characterized by the equation 
\begin{eqnarray}
\Qhat V(z) &=& \del U(z) \comma \label{defintU} 
\end{eqnarray}
so that $\int [dz] V(z)$ is annihilated by $\Qhat$. 
 Besides the freedom of adding 
a $\Qhat$-exact term 
 $\Qhat \Sigma$, $V$ has the ambiguity inherited from 
 the change $U \rightarrow U +\Qhat \Omega$, which amounts to adding 
 the total derivative $\del \Omega$. 
\subsubsection{Construction by homological perturbation}
One way to construct $V$ is to make use of the scheme of 
homological perturbation, just as in the case of the unintegrated vertex. 
Decomposing (\ref{defintU}) 
 according to the degree, we get the master equation 
\begin{eqnarray}
&& V = \sum_{n=n_{min}}^5 V_n \comma \\
&& \del U_n -QV_n - d_1 V_{n-1} -d_2 V_{n-2} = \delta V_{n+1} \comma 
\label{intmastereq}
\end{eqnarray}
where $n_{min}$ is the lowest degree allowed for $V_n$. Contrary to 
the case of $U$, $n_{min}$ can be as low as $-1$ since $V$ is of dimension
 1 and a structure like $b^\mu X_\mu$ of degree $-1$ is possible. 
As it will be clear  shortly, there are two types of solutions, related 
 by a $\Qhat$-exact term,  with either $n_{min}=-1$ or $n_{min}=0$. 

Let us briefly discuss the $n_{min}=0$ case, which gives a 
direct extension of the conventional
 Berkovits vertex \cite{Berk0001,Siegel86}
given by 
\begin{eqnarray}
V^B_0 &=& \del \theta^\al A_\al + \Pi^\mu B_\mu + d_\al W^\al 
 + \half \Llam^{\mu\nu}F_{\mu\nu} \comma   \label{VBzero}
\end{eqnarray}
where $\Llam^{\mu\nu}$ is the Lorentz generator for the $(\lam, \omega)$ 
 sector:
\begin{eqnarray}
\Llam^{\mu\nu} &=& -\half (\omega \ga^{\mu\nu} \lam) \period \label{Lorlam}
\end{eqnarray}
With the PS constraints, this vertex satisfies the 
$n=0$ part of the master equation (\ref{intmastereq})
\begin{eqnarray}
\del U_0 -QV_0 &=& \delta V_1 \comma \label{intdegzero}
\end{eqnarray}
with $V_1=0$.  When the PS constraints are removed, however, 
this no longer holds and one gets $\del U_0 -QV^B_0 = i\Lam^\mu 
(\omega \del_\mu W)$. The apparent problem is that this cannot possibly 
 be written as $\delta V_1$ since the fact that $Q\delta V^B_0\ne 0$
 contradicts the consistency equation obtained upon acting $\delta$ 
 on (\ref{intdegzero}). The solution to this problem  turned out to be 
that one can modify $V^B_0\rightarrow V_0$ in such a way that $V_0$ satisfies 
 both $\delta V_0=0$ and (\ref{intdegzero}) simultaneously, with an appropriate
 $V_1$. 

Once this is achieved, the rest of the construction is straightforward 
thanks to the triviality of $\delta$-homology proved in Appendix B:
 By using the the set of equations (\ref{delsq}) $\sim$ (\ref{dtwosq}) 
 following from the nilpotency of $\Qhat$, it is easy to show that 
the LHS of (\ref{intmastereq}) is annihilated by $\delta$ and hence 
 can be written as $\delta V_{n+1}$, from which one can read off $V_{n+1}$. 
However, we shall not give the details of this procedure because 
there exists a  conceptually superior way to 
 construct $V$ with  much less effort, to which we now turn. 
\subsubsection{\lq\lq $b$-ghost", Virasoro operator and construction of $V$}
It is well-known that the action of the worldsheet derivative $\del$
 is implemented by that of $L_{-1} = \int [dz] T(z)$, where $T(z)$ 
 is the energy-momentum tensor,  given in our case by
 (\ref{emtensor}). 
Now suppose we can find the \lq\lq $b$-ghost" field $B(z)$ such that 
\begin{eqnarray}
T(z) &=& \Qhat B(z) \equiv \acom{\Qhat}{B(z)} \period \label{TQB}
\end{eqnarray}
Then, if we define  $\calB \equiv \int[dz] B(z)$, we have 
$L_{-1} = \Qhat \calB $ and hence acting on the unintegrated vertex $U$ 
one gets 
\begin{eqnarray}
\del U &=& (\Qhat \calB) U = \Qhat (\calB U) + \calB(\Qhat U) = \Qhat (\calB U)
\comma  \label{delU}
\end{eqnarray}
which is nothing but the defining equation for the integrated
 vertex $V$.  Therefore we can construct  $V$ simply as 
\begin{eqnarray}
V &=& \calB U \period \label{VBU}
\end{eqnarray}

A gratifying fact is that one can find  such $B(z)$  in a 
 remarkably simple form. It consists of degree $-1$ and $0$ pieces and 
 is given by
\begin{eqnarray}
B(z) &=& B_{-1}(z) + B_0(z) \comma \\
B_{-1} &=& -b^\mu \Pi_\mu \comma \qquad B_0 = -\omega_\al \del\theta^\al
\period \label{Bghost}
\end{eqnarray}
Notice that $B(z)B(w) = -b_I(z) b_J(w) \calNbar^\mu_I \calNbar^\mu_J 
/(z-w)^2 =0$, which is a 
 desired property. 

Let us demonstrate that indeed $\Qhat B(z) = (\delta + Q + d_1 +d_2)B(z)
 =T(z)$. First it is almost trivial to show that 
$Q B_0(z) = \del\theta^\al d_\al -\omega_\al \del \lam^\al$ and 
 $d_1 B_{-1}(z) = -\half \Pi^\mu \Pi_\mu -b_\mu \del c^\mu$, 
 so that
\begin{eqnarray}
T(z) &=& Q B_0(z) +d_1 B_{-1}(z) \period 
\end{eqnarray}
Next, it is just as easy to prove $\delta B_0(z) + QB_{-1}(z) =0$ 
 and $\delta B_{-1}(z) =0$. The remaining OPE's require 
non-trivial calculations.  They can be  performed with the help of 
 $U(5)$ decompositions of certain quantities, details of which are 
 not particularly illuminating and hence are omitted.
 The results, however, are extremely simple and 
 we get $d_1 B_0(z)+d_2 B_{-1}(z) =0$ and $d_2B_0(z) =0$. 
Altogether the proof of  $\Qhat B(z) =T(z)$ is completed. 

Now the complete integrated vertex $V$ is  succinctly given by 
 $V=(\calB_{-1} + \calB_0) U$  and  can be easily computed explicitly. 
The structures up to degree 0 turned out to be  particularly 
 intriguing. A simple calculation gives 
\begin{eqnarray}
V &=& (\calB_{-1} + \calB_0) U = b^\mu \lam^\al \del_\mu A_\al 
 + \del\theta^\al A_\al + \half \Pi^\mu B_\mu -b^\mu c^\nu \Fmn\nn\\
&&  + \calNtil^{\mu\nu} \Pi_\mu B_\nu 
+ c^\mu b^\nu  \del_\mu B_\nu + V_{n\ge 1} \period
\end{eqnarray}
where we defined $\calNtil^{\mu\nu} \equiv \half (\calNbar^\mu_I \calN^\nu_I
 -\calNbar^\nu_I \calN^\mu_I)$. 
Evidently, $V$ starts out from dergree $-1$ and its structure at degree 
0 is, surprisingly, rather different from the Berkovits vertex $V^B_0$:
 Among other things,  $d_\al W^\al 
 + \half \Llam^{\mu\nu}F_{\mu\nu}$ part is missing and 
 the coefficient of the term $\Pi^\mu B_\mu$ is only a half. 

The solution to this puzzle is provided by the freedom of 
adding $\Qhat$-exact terms. First one can show  that the degree $-1$ part of 
 $V$ above can be rewritten exactly as $b^\mu \lam^\al \del_\mu A_\al
= -\delta (\omega W) -Q(b^\mu B_\mu)$. This means that if we add
 to $V$ the terms $\Qhat((\omega W) +(b^\mu B_\mu))$, the resultant vertex 
 starts from degree 0. Now a rather remarkable fact is that $Q(\omega W)$ 
 contained in $\Qhat(\omega W)$ gives 
\begin{eqnarray}
Q(\omega W) &=& d_\al W^\al  + \half \Llam^{\mu\nu}F_{\mu\nu} \comma 
\end{eqnarray}
which is exactly the structure we are looking for. Moreover the 
missing half of $\Pi^\mu B_\mu$ is supplied by $d_1(b^\mu B_\mu)$ 
 in $\Qhat(b^\mu B_\mu)$. In this way the alternative vertex 
becomes 
\begin{eqnarray}
\Vtil &=& V + \Qhat (\omega W + b^\mu B_\mu) \nn\\
&=& V^B_0 -b^\mu c^\nu \Fmn
+\half \del \calNtil^{\mu\nu} \Fmn +\Vtil_{n\ge 1} \comma 
\end{eqnarray}
which indeed contains precisely the Berkovits vertex $V^B_0$. 
As expected,  this vertex coincides with the one constructed by 
the homological perturbation technique described previously. 

Besides being extremely useful for the construction of $V$, 
the existence of a simple form of \lq\lq $b$-ghost" we have uncovered
 would  have  far-reaching consequences. The relation to the Virasoro
 generator, which hitherto has been rather elusive in PS formalism, 
is now clearly understood and the relation $T=\acom{\Qhat}{B} $ is known 
 to be of prime importance for no-ghost theorem and loop calculations 
\cite{FreemanOlive}. Study of its significance in our formalism 
 is now under investigation and will be reported elsewhere \cite{AK2}. 
\section{Summary and Discussions}
In this paper, we have presented a new extension of the Berkovits' 
 pure spinor formalism for superstring, in which pure spinor constraints 
 are removed in a rather natural and efficient way with a minimum number
 of ghost fields. As a summary, it should be helpful to make a
list of characteristic features of our formalism. 
\begin{itemize}
	\item With all the constraints removed, the question of how to 
actually treat the constraints in quantization procedure has been exorcized. 
	\item The Hilbert space structure is clarified and it is possible 
 to realize the peculiar hermiticity property of $\lam^\al$ 
with the aid of a modified Fock space inner product. 
	\item PS constraints are precisely captured without redundancy 
 and this led to the new simple first class algebra. 
	\item In constructing the nilpotent BRST-like charge, a
 minimum number of additional ghosts $(c_I, b_I)_{I=1\sim 5}$ are required. 
	\item Everything fits nicely into the scheme of homological perturbation 
 theory and the proof of the equivalence of cohomologies as well as the 
construction of massless vertex operators are achieved in a systematic manner. 
	\item A simple composite \lq\lq $b$-ghost" field $B(z)$ is constructed 
 which realizes the fundamental equation $T(z) = \acom{\Qhat}{B(z)} $
 and  the relation to the Virasoro operator is thereby clarified. 
As an application, construction of the integrated vertex is made extremely 
 efficient. 
\end{itemize}

Obviously there are many remaining problems to be investigated and clarified. 
(i)\ First and foremost, although the Lorentz invariance of the cohomology is 
 assured, we would like to further extend our formalism, by introducing 
 additional ghosts, and achieve manifest Lorentz covariance. 
(ii)\  The calculation of scattering amplitudes 
 should be performed and the rules of computation should be derived. 
(iii)\ Further consequences of the important relation 
$T(z) = \acom{\Qhat}{B(z)} $ should be investigated. 
(iv)\ Relation to the RNS formalism needs to be clarified. 
(v)\ There should be no problem in applying our formalism to a superparticle 
 case. Application to a supermembrane, on the other hand,  is expected to 
be non-trivial and interesting. (vi)\ Other obvious problems are the extension 
 to the case of curved background and description 
of D-branes in our formalism. 
These and related matters are under investigation and results will be 
 reported elswhere \cite{AK2}. 
\par\bigskip\noindent
{\large\bf Acknowledgment}\par\smallskip\noindent
We are grateful to  M.~Naka for his interest and collaboration 
 at an early stage of this work. Y.K. thanks M.~Kato for a useful
 discussion. 
The research of Y.K. is supported in part by the 
 Grant-in-Aid for Scientific Research (B) 
No.~12440060 from the Japan  Ministry of Education,  Science
 and Culture. 

\appendix
\setcounter{equation}{0}
\renewcommand{\theequation}{A.\arabic{equation}}
\section*{Appendix A: \ \
Conventions and Useful Formulas}
In this appendix, we collect our conventions and some 
 useful formulas employed in the text. 
\subsection*{A.1.\ \ Spinors and $\Ga$-matrices in real basis}
$32\times 32$ $SO(9,1)$ Gamma matrices are denoted by $\Ga^\mu, 
 (\mu=0,1,\ldots, 9)$ and obey the Clifford algebra 
 $\acom{\Ga^\mu}{\Ga^\nu} =2\eta^{\mu\nu}$. Our metric convention is
 $\eta^{\mu\nu} = (-,+,+,\ldots, +)$. The 10-dimensional chirality 
 operator is taken to be $\Gabar_{10}=\Ga^0\Ga^1\cdots \Ga^9$ and 
 it satisfies  $\Gabar_{10}^2=1$. 

In the Majorana or real basis (R-basis
 for short), $\Ga^\mu$ are all real and unitary. Within the R-basis, 
 we define the Weyl basis to be the one in which $\Gabar_{10} =${\rm diag}\
 $(1_{16}, -1_{16})$, where $1_{16}$ is the $16\times 16$ unit matrix. 
In this basis, a general 32-component spinor $\Lam$ is written as 
$\Lam = \vecii{\lam^\al}{\lam_{\al}} $, where $\lam^\al$ 
and $\lam_\al$ are chiral and anti-chiral respectively, 
with $\al=1 \sim 16$. Correspondingly, $\Ga^\mu$, which flips chirality, 
 takes the structure 
\begin{eqnarray}
\Ga^\mu &=& \matrixii{0}{(\ga^\mu)^{\al\be}}{(\ga^\mu)_{\al\be}}{0} 
\comma \label{GainWeyl} 
\end{eqnarray}
where the $16\times 16$ $\ga$-matrices $(\ga^\mu)^{\al\be}$ and 
 $(\ga^\mu)_{\al\be}$ are real symmetric and satisfy 
\begin{eqnarray}
(\ga^\mu)_{\al\be} (\ga^\nu)^{\be\ga} 
+ (\ga^\nu)_{\al\be} (\ga^\mu)^{\be\ga} =2\eta^{\mu\nu}\delta_\al^\ga 
\period \label{gaClif}
\end{eqnarray}
In terms of $\ga^\mu$, an often used Fierz identity is expressed 
as $(\ga_{\mu})_{\al\be} (\ga^\mu)_{\ga\delta} 
 + (\mbox{cyclic in $\al, \be, \ga$}) =0$. 

Anti-symmetrized products of $\ga^\mu$ are defined in the following 
way. $(\ga^{\mu\nu})^\al{}_\be$ and $(\ga^{\mu\nu})_\be{}^\al$ are 
 defined as 
\begin{eqnarray}
(\ga^{\mu\nu})^\al{}_\be &=& \half \Bigl( (\ga^\mu)^{\al\ga}
 (\ga^\nu)_{\ga\be} 
 -(\ga^\nu)^{\al\ga} (\ga^\mu)_{\ga\be} \Bigr) \comma \\
(\ga^{\mu\nu})_\be{}^\al  &=& \half \Bigl( (\ga^\mu)_{\be\ga}
  (\ga^\nu)^{\ga\al}  
 -(\ga^\nu)_{\be\ga} (\ga^\mu)^{\ga\al} \Bigr) \comma 
\end{eqnarray}
and they are related by $(\ga^{\mu\nu})^\al{}_\be = -(\ga^{\mu\nu})_\be{}^\al$.
$\ga^{\mu_1\mu_2\ldots \mu_k}$'s are similarly defined. Anti-symmetric 
products with  odd number of $\ga$'s have definite symmtery properties. 
$\ga^\mu$ and $\ga^{\mu_1 \ldots \mu_5}$ are symmetric, while 
 $\ga^{\mu_1\mu_2\mu_3}$ is anti-symmetric. 
$(\ga^{\mu_1 \ldots \mu_5})^{\al\be}$ is self-dual in the sense 
$(\ga^{\mu_1 \ldots \mu_5})^{\al\be} =(1/5!) \ep^{\mu_1\ldots \mu_5 \nu_1
 \ldots \nu_5} (\ga_{\nu_1 \ldots \nu_5})^{\al\be}$, where 
 $\ep^{012\ldots 9} \equiv 1$. Similarly, 
$(\ga^{\mu_1 \ldots \mu_5})_{\al\be}$ is anti-self-dual.

Finally, we note that Lorentz covarint spinor bilinears of 
$\Lam =\vecii{\lam^\al}{\lam_{\al}} $
and  $\Psi= \vecii{\psi^\al}{\psi_\al} $ can be constructed, 
 without the use of  complex conjugation,   as
\begin{eqnarray}
\Lam^T C \Psi &=& \lam^\al \psi_\al -\lam_\al \psi^\al \comma \\
\Lam^T C \Ga^\mu \Psi &=& \lam^\al \ga^\mu_{\al\be} \psi^\be 
 -\lam_\al (\ga^\mu)^{\al\be} \psi_\be \comma \\
\Lam^T C \Ga^{\mu\nu} \Psi &=& \lam^\al (\ga^{\mu\nu})_\al{}^\be 
 \psi_\be -\lam_\al (\ga^{\mu\nu})^\al{}_\be \psi^\be \comma \quad 
etc \comma 
\end{eqnarray}
where $C$ is the charge conjugation matrix. 
\subsection*{A.2. \ \ {\boldmath $U(5)$} basis}
It is well-known that the spinor representations for $SO(9,1)$ 
 and $SO(10)$ can be conveniently 
 constructed with the use of 5 pairs of 
 fermionic oscillators $(b_I, b_I^\dagger)$ satisfying the anti-commutation
 relations $\acom{b_I}{b_J^\dagger} = \delta_{IJ}$. States are built upon 
 the oscillator vacuum, to be  denoted by $\ket{+} $, annihilated by 
all the $b_I$'s. 
It is clear that the anti-commutation relations are invariant 
 under the action of $U(5)$, where $b_I$ and $b_I^\dagger$ transform 
 repsectively as $5$ and $\bar{5}$ (or $\bar{5}$ and $5$, depending on
 one's convention). 
$\Ga^\mu$ matrices can then be regarded as linear operators in this 
 Fock space  and in the case of $SO(10)$ they are identified as 
\begin{eqnarray}
\Ga^{2I} &=& {1\over i} (b_I-b_I^\dagger)\comma 
 \quad \Ga^{2I-1} = b_I +b_I^\dagger \comma \quad I=1\sim 5 \period 
\end{eqnarray}
Since the $SO(9,1)$ case is easily recovered  by setting 
 $\Ga^0=i\Ga^{10} =b_0-b_0^\dagger$, where $(b_0, b_0^\dagger)
 \equiv (b_5, b_5^\dagger)$,  
 we will use $SO(10)$ notations. 

The states built upon $\ket{+} $ and their conjugates are defined  as 
\begin{eqnarray}
\ket{I_1I_2 \ldots I_k} &\equiv & b_{I_1}^\dagger \cdots 
 b_{I_k}^\dagger \ket{+} \comma  \label{U5ket} \\
\bra{I_1I_2 \ldots I_k} &\equiv & \bra{+} b_{I_k} b_{I_{k-1}}
 \cdots b_{I_1} \period \label{U5bra} 
\end{eqnarray}
Further, we define
\begin{eqnarray}
\ket{-} &\equiv & b_1^\dagger b_2^\dagger \ldots b_5^\dagger \ket{+}
 = {1\over 5!} \ep_{I_1I_2 \ldots I_5} b_{I_1}^\dagger b_{I_2}^\dagger
 \cdots b_{I_5}^\dagger \ket{+} \comma \label{minusket} \\
\ket{\Itil_1 \ldots \Itil_k} &\equiv & {1\over (5-k)!} 
 \ep_{I_1 \ldots I_k J_{k+1} \ldots J_5} \ket{J_{k+1} \ldots J_5} 
 \comma \label{tildeket}
\end{eqnarray}
and their corresponding conjugates, where $\ep_{12345} \equiv 1$. 
These states satisfy the orthonormality relations 
\begin{eqnarray}
\bk{+}{+} &=& \bk{-}{-} =1 \comma \\
 \bk{I_1 \ldots I_k}{J_1 \ldots J_k} &=& 
\bk{\Itil_1 \ldots \Itil_k}{\Jtil_1 \ldots \Jtil_k}
 = \delta^{I_1 \ldots I_k}_{J_1\ldots J_k} \period 
\end{eqnarray}
In this basis, chiral and anti-chiral spinors can be written as 
\begin{eqnarray}
 \mbox{chiral:} \qquad \ket{\lam} 
 &=& \lam_+\ket{+} + \half \lam_{IJ} \ket{IJ} + \lam_\Itil \ket{\Itil} 
 \comma \label{U5chiral} \\
 \mbox{anti-chiral:} \qquad \ket{\psi} 
 &=& \psi_-\ket{-} + \half \psi_{\Itil \Jtil} \ket{\Itil \Jtil} 
+ \psi_I \ket{I} 
 \period \label{U5achiral}
\end{eqnarray}
We write the general components of chiral and anti-chiral spinors 
 as $\lam_A =\bk{A}{\lam} $ and $\psi_\Abar=\bk{\Abar}{\psi} $. 

The charge conjugation matrix in this basis is given 
 by
\begin{eqnarray}
C &=&  -i\Ga^2\Ga^4 \Ga^6\Ga^8\Ga^{10} =-\Ga^0 \Ga^2\Ga^4 \Ga^6\Ga^8 \period
\label{U5C}
\end{eqnarray}
Its action on the states is  
\begin{eqnarray}
C\ket{+} &=& \ket{-} \comma \quad C\ket{IJ} = -\ket{\Itil \Jtil}
 \comma \quad C \ket{\Itil} = \ket{I} \comma \\
C\ket{-} &=& -\ket{+} \comma \quad C\ket{\Itil \Jtil} = \ket{IJ} 
\comma \quad C\ket{I} = -\ket{\Itil} \period \label{Caction}
\end{eqnarray}

\setcounter{equation}{0}
\renewcommand{\theequation}{B.\arabic{equation}}
\section*{Appendix B: \ \
Triviality of $\delta$-homology for degree $\ge 1$}
In this appendix, we give a proof of 
 the triviality of the $\delta$-homology
 for degree $\ge 1$ in the general space of operators consisting 
 of arbitrary number of $x^\mu, \theta^\al, \lam^\al, \omega_\al,
c_I, b_I$ and their worldsheet derivatives.

The action of $\delta=-i\int[dz]\Phi_I(z)b_I(z)$ in the sense 
 of operator product 
can be represented more explicitly by the following three types of 
 variational operations:
\begin{eqnarray}
\delta &=& \delta_1 + \delta_0 \comma \\
\delta_0 &=& \delta_{0a} + \delta_{0b} \comma \\
\delta_1 &=& -i\int[dz] \Phi_I(z) {\delta \over \delta c_I(z)} \comma \\
\delta_{0a} &=& -i \int[dz] b_I(z) G_{I\al}(z) {\delta \over 
\delta \omega_\al(z)} \comma \\
\delta_{0b} &=& -i \int[dz] G_{I\al}(z) {\delta \over \delta c_I(z)}
{\delta \over \delta \omega_\al(z)} \period
\end{eqnarray}
Here the variational derivative $\delta/\delta a(z)$ is defined 
 as $(\delta/\delta a(z)) a(w) = (z-w)^{-1}$ and 
 $G_{I\al}(z)$ is the residue of the simple pole in the 
  OPE $\Phi_I(z) \omega_\al(w)  = G_{I\al}(z)/(z-w)$. 
$\delta_1$ 
and $\delta_{0a}$ effect the single contraction with $b_I(z)$ and $\Phi_I(z)$
 respectively and  $\delta_{0b}$ represents the double contraction. 
The subscripts 0 and 1 signify the number of $\Phi_I$, to be referred 
 to as ``$\Phi$-level",  increased by 
 the operation\footnote{Alternatively, ``$\Phi$-level" is the same 
as the number of $\lam_\Itil$.}. Actually, when $\delta_1$ 
acts on $\del^m c_I$ it produces 
 $\del^m\Phi_I$, but the derivatives acting on $\Phi$ will be irrelevant 
 in the counting of $\Phi$-level. 
Clearly $\delta_1, \delta_{0a}$ and $\delta_{0b}$ 
anticommute with each other and with themselves. 

In particular, $\delta_1^2=0$ and hence one can define $\delta_1$-homology. 
We shall now show that the demonstration of the triviality of 
$\delta$-homology can be reduced to that of $\delta_1$-homology. 
Since any operator not containing $c_I$ (or its worldsheet derivative)
 is annihilated  by $\delta_1$, $\delta_1$-homology is 
obviously non-trivial at degree 0. Hence the following argument applies
to operators with degree $\ge 1$. 

Assume that $\delta_1$-homology has been shown to be trivial and 
let $\calH_N$ be the space of operators 
at $\Phi$-level $N$ or lower, where $N$ can be arbitrarily large. 
Then, any operator $A\in \calH_N$ can be written as 
\begin{eqnarray}
 A &=& \sum_{n=0}^N A_n \comma
\end{eqnarray}
where $A_n$ is at $\Phi$-level $n$. Now suppose $\delta A=0$. Decomposing
 this equation according to the $\Phi$-level, we have
\begin{eqnarray}
0 &=& \delta_0 A_n + \delta_1 A_{n-1}\comma \qquad 0 \le n \le N+1 \comma 
\end{eqnarray}
with $A_n \equiv 0$ for $n < 0$ and $n >N$. At the highest $\Phi$-level, 
we have $\delta_1 A_N=0$ and hence by the triviality assumption
$A_N$ must be $\delta_1$-exact, namely it can be written as 
 $A_N =\delta_1 B_{N-1}$. Using this result, the closedness relation 
at $\Phi$-level $N-1$ becomes 
\begin{eqnarray}
0 &=& \delta_0 A_N + \delta_1 A_{N-1} =\delta_0 \delta_1 B_{N-1} + 
\delta_1 A_{N-1} \nn\\
&=& \delta_1 (A_{N-1} -\delta_0 B_{N-1}) \comma 
\end{eqnarray}
which can be solved as $A_{N-1} = \delta_0 B_{N-1} + \delta_1 B_{N-2}$. 
Continuing in this fashion, we can easily obtain the result 
$A= \delta B$, where $B=\sum_{n=0}^{N-1}B_n$. This proves the triviality 
 of the $\delta$-homology. 

Thus it suffices to prove the triviality of the $\delta_1$-homology. 
Let $\calH_{M,N}$ be the space of operators where the largest number 
 of $\del$ on $c_I$ is $M$ and the largest number of such factor, 
$\del^M c_I$, is $N$. $M$ and $N$ can be arbitrarily large. 
Assume that we have proved the 
 triviality of $\delta_1$ up to  $\calH_{M-1, *}$ where $*$ is 
 arbitrary. A general operator 
 $A_M$ in $\calH_{M,N}$ can then be written as
\begin{eqnarray}
A_M &=& \sum_{n=0}^N A_{M,n} \comma \\
A_{M,n} &=& 
\del^M c_{I_1} \del^M c_{I_2} \cdots \del^M c_{I_n}
\Atil_{I_1 I_2 \ldots I_n} \comma 
\end{eqnarray}
where $\Atil_{I_1 I_2 \ldots I_n}$ as well as $A_{M,0}$ are 
in $\calH_{M-1, *}$. 
Now suppose 
$i\delta_1 A_M=0$. Since the only term in $i\delta_1 A_M$ containing the 
 largest number, 
$N$, of  $\del^M c_I$'s is 
$(-1)^N \del^M c_{I_1} \del^M c_{I_2} \cdots \del^M c_{I_N}
\delta_1 \Atil_{I_1 I_2 \ldots I_N}$, it must vanish separately. 
This means $\delta_1 \Atil_{I_1 I_2 \ldots I_N} =0$. From our assumption, 
this can be solved as 
\begin{eqnarray}
\Atil_{I_1 I_2 \ldots I_N} &=& \delta_1 \Btil_{I_1 I_2 \ldots I_N} \period
\end{eqnarray}
Therefore we can write 
\begin{eqnarray}
A_{M,N} &=& 
\del^M c_{I_1} \del^M c_{I_2} \cdots \del^M c_{I_N} \delta_1 
\Btil_{I_1 I_2 \ldots I_N} \nn\\
&=&  \delta_1 ((-1)^N \del^M c_{I_1} \del^M c_{I_2} \cdots \del^M c_{I_N} 
\Btil_{I_1 I_2 \ldots I_N} ) \nn\\
&& \quad  -(-1)^N N \del^M \Phi_{I_1} 
\del^M c_{I_2} \cdots \del^M c_{I_N} 
\Btil_{I_1 I_2 \ldots I_N} \period
\end{eqnarray}
The term in the second line is $\delta_1$ exact (and hence automatically 
 closed) while the term in the third line has only $N-1$ $\del^M c_I$'s and 
can be absorbed into $A_{M,N-1}$. This means that effectively 
 $A_{M,N}$ can be removed from our analysis. 
Repeating this argument, we can remove all the $\del^M c_I$ factors. 
We will then be left with operators belonging to $\calH_{M-1,*}$, 
for which the triviality of $\delta_1$ holds by assumption.

In this way, by mathematical induction, 
the problem is reduced to the case of the space $\calH_{0,*}$, 
\ie, for operators without any $\del$'s on $c_I$'s. The proof in this case,
however, 
 is formally the same as the simple situation  already described 
 in the main text: The action of $\delta$ in that case is operationally 
 identical to that of $\delta_1$ in the present more general situation
 and hence $\delta_1 A=0$ implies $A=\delta_1 B$,
 due essentially to the algebraic independence of $\Phi_I$. 
This completes the proof. 

\end{document}